\documentclass[a4paper,10pt]{elsarticle}

\usepackage{lineno}
\usepackage[margin=0.7in]{geometry}
\usepackage{appendix, etex, lmodern, longtable, parskip, ragged2e}
\usepackage{adjustbox, enumitem, etoolbox, subscript, tabularx, tabu}
\usepackage{optidef}
\usepackage{amsmath,amsfonts,amssymb,amsthm, bm}
\usepackage{adjustbox}
\usepackage{mathtools}
\usepackage[utf8]{inputenc}
\usepackage[dotinlabels]{titletoc}
\usepackage{multirow}
\usepackage{longtable}
\usepackage[english]{babel}
\usepackage{hyperref}
\usepackage[table]{xcolor,xcolor}
\usepackage{color,soul}
\usepackage{graphicx}
\usepackage{caption}
\usepackage{subcaption}
\usepackage{pdflscape}
\usepackage{tabularx}
\usepackage{rotating}
\usepackage{amssymb,mathtools}
\usepackage{float}
\usepackage{booktabs}
\usepackage{ltablex}
\usepackage[table]{xcolor}
\definecolor{mycolor}{rgb}{0.9, 0.9, 0.9} 
\usepackage{threeparttable}

\journal{}

\begin{document}
\begin{frontmatter}

\title{From Survey Personas to LLM Agents: A Generative Agent-based Simulation of Mobility Policy Preference Dynamics}

\author{Ali Torkayesh  \fnref{myfootnote1}}
\cortext[cor1]{Corresponding author}
\ead{torkayesh@sustainability.rwth-aachen.de}
\address[myfootnote1]{Economics of Sustainability, RWTH Aachen University, Germany}

\author{Julia Offermann\fnref{myfootnote2}}
\address[myfootnote2]{Chair of Communication Science, RWTH Aachen University, Germany}

\author{Regina Gimpel \fnref{myfootnote3}} \address[myfootnote3]{Risk Perception and Communication, RWTH Aachen University, Germany}

\author{Linda Engelmann \fnref{myfootnote2}} 

\author{Katrin Arning \fnref{myfootnote3}}

\author{Martina Ziefle \fnref{myfootnote2}}

\author{Sandra Venghaus \fnref{myfootnote1}}

\begin{abstract}

Large language models (LLMs) have been increasingly used to simulate socially complex and interaction-driven tasks. However, most existing studies rely on hand-crafted personas. Since persona design strongly shapes how agents interpret context and make decisions, developing empirically grounded agent profiles is a significant aspect in this underexplored research area. To address this limitation, we propose a survey-grounded generative agent-based modeling (GABM) simulation framework that translates real survey respondents into generative LLM agents. The main objective of our framework is to demonstrate how careful persona design enables realistic simulation of decision-making using LLMs for facilitating behavioral experiments. We illustrate the framework's applicability through a case study of mobility policy preference dynamics in Germany, focusing on public support for phasing out new internal combustion engine vehicles, which is part of the European Union's net-zero target. Our benchmark is based on 514 survey respondents, each translated into a natural-language persona, grounded in demographic characteristics, political orientation, mobility behavior, fuel experience, and climate-related attitudes. The simulation goal is to examine how support evolves over time, how agents switch positions across rounds, and how responses differ across survey-grounded personas under changing social and policy contexts.

\end{abstract}


\end{frontmatter}


\begin{figure*}[h]
  \includegraphics[width=\linewidth]{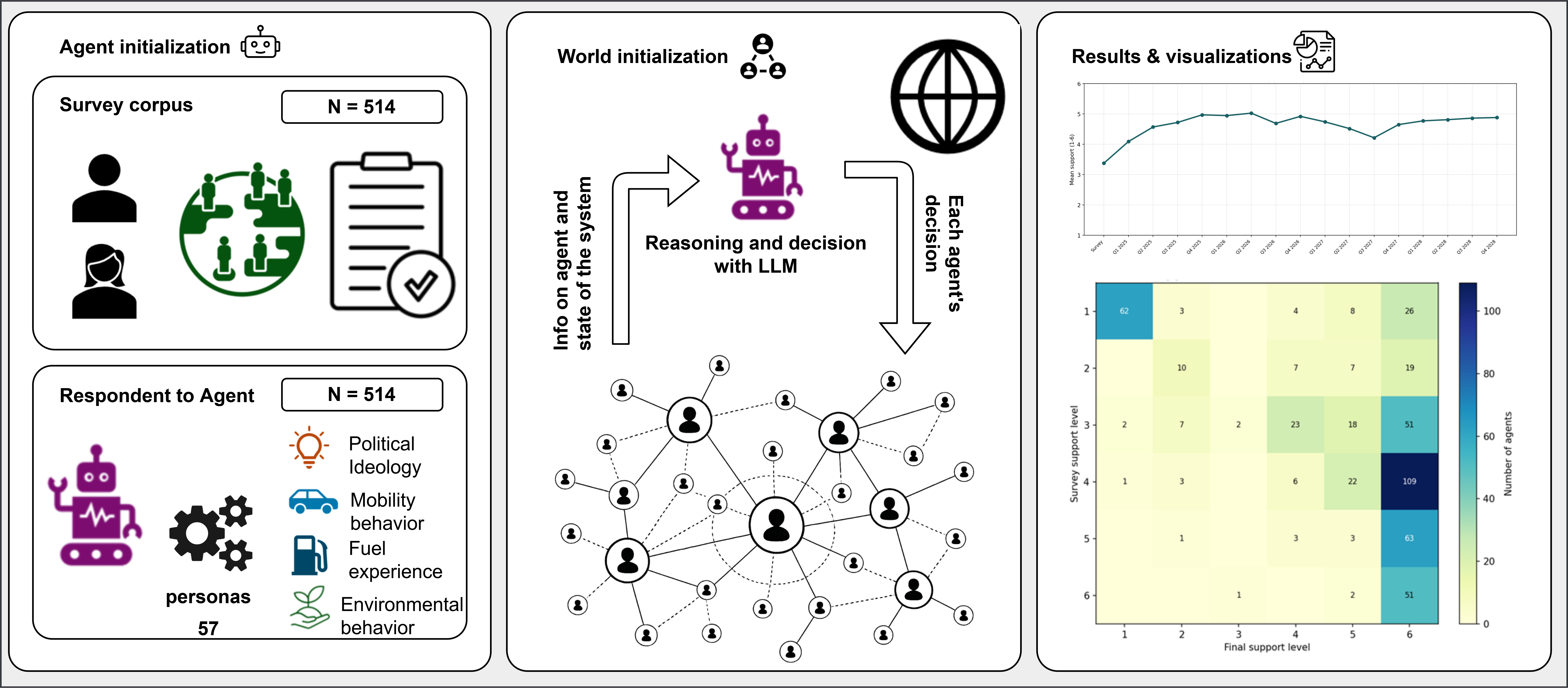} \hfill
  \caption {Overview of the LLM-based GABM simulation framework.}
  \label{Fig:Framework}
\end{figure*}

\section{Introduction}
The European Union has expressed a strong commitment to phasing out new internal combustion (ICE) vehicles through its Fit for 55 package as part of its broader net-zero transition \citep{Regul2023}. However, public responses to such major policy change do not emerge in isolation \citep{hoppe2023public}. In such social and political complex systems, public responses are shaped through interaction, feedback, and changing social, economic, environmental, and political contexts \citep{kumar2026simulation,ahrens2024impact,moussaid2013social}. In this context, surveys are the major tools used to obtain empirical evidence, which are well-suited to measuring attitudes at specific points in time but less suited to modeling how those attitudes evolve under exogenous shocks and unfolding events.

The recent literature shows a growing body of studies interested in exploring how large language models (LLMs) as generative agents could be used to simulate human behavior in socially grounded tasks \citep{guo2024large,mou2024individual,sreedhar2025simulating}. This development has opened new opportunities for computational social science, including emerging applications in the climate change domain \citep{nabavi2026using}. LLMs enable generative agents, built on the survey respondents, to reason over natural-language personas, social context, and unfolding events in order to explore dynamic preference formation beyond static observational data. In order to move beyond manually hand-crafted personas or stylized settings for LLM agents, we propose constructing LLM agents from an empirically grounded agent population derived from survey data. In this context, two key gaps exist in the literature: i) how to construct LLM agents from empirically grounded respondent populations rather than hand-crafted personas, and (ii) how to use these agents to study dynamic policy preference formation under changing social context and external events.

In this study, we aim to address these critical gaps and present the following contributions. First, we introduce a survey-grounded generative agent-based modeling (GABM) simulation framework that transforms survey respondents into LLM agents, and represents them through natural-language personas capturing demographic background, political orientation, mobility behavior, fuel experience, and climate-related attitudes (Figure \ref{Fig:Framework}). Second, we apply this framework to a significant and timely important mobility policy preference dynamics in Germany, focusing on public support for phasing out new ICE vehicles. Finally, we provide an empirical analysis of how survey-grounded LLM agents respond to changing social context and external developments. Understanding responses over time facilitates examining temporal support trajectories, preference switching, and heterogeneity across agent profiles.

\section{Related Works}
With recent advances in the LLM research domain, there is a notable growth in the body of studies using LLMs for agent-based simulation through modeling socially complex and interaction-driven environments within natural-language reasoning \citep{kwon2025evaluating,huang2024personality,wang2024rolellm,jia2024can}. Key advancements have been made in the use of LLMs for simulating human behavior by emulating human traits and their reflection in the decision-making process \citep{wang2024rolellm,kwon2025evaluating}. In this context, decision-making represents a particularly challenging form of human behavior for LLM agents, as it requires not only strong reasoning but also sensitivity to the conditions under which decisions are made \citep{tamkin2021understanding,huang2024personality}. In real-world settings, decision-making depends not only on reasoning itself, but also on the social, environmental, and cognitive context in which a choice is formed \citep{huang2024personality}.

This challenge has motivated a recent work on GABM, which combines explicit simulation environments with LLM agents to interpret context and generate decisions over time \citep{ghaffarzadegan2024generative}. GABM framework developed by \citet{ghaffarzadegan2024generative} offers a promising framework to integrate natural-language reasoning into agent-based systems in order to explore dynamic social processes that unfold through interaction, feedback, and exogenous shocks. Despite the growing use of GABM across different application domains, a key question remains in socially grounded decision-making tasks: how can LLM agents be constructed in ways that more closely reflect empirically observed human heterogeneity?

To address this gap, we present a survey-grounded GABM framework that constructs rich personas for LLM agents to better reflect real-world decision-makers. Using the case of the ICE phase-out in the European Union, we evaluate the behavior of survey respondents as LLM agents over time. 

\section{Methodology}
This section introduces the simulation framework based on LLM agents with personas derived from respondents of a survey (Figure \ref{Fig:Framework}). In Section \ref{Data}, we first describe the survey data and the variables (personas) used to build the agent profiles. Next, in Section \ref{SimFramework}, we present the simulation framework, including persona construction, social context, and event-driven interaction. Finally, in Section \ref{Experiment}, we outline the experimental setup used to evaluate the model under baseline and event-driven conditions.

\subsection{Data}
\label{Data}
Survey design plays the key role in deriving structured data needed to construct empirically grounded agent personas. Rather than treating the survey as a generic instrument, we designed the survey based on the Integrative Public-Policy-Acceptance (IPAC) framework \citep{grelle2024and}. The IPAC provides a
comprehensive structure for capturing person-, context-, policy-specific determinants, and policy acceptance \citep{grelle2024and}. An online survey for German laypeople over the age of 18 was conducted to assess the adaptation of the IPAC for the phase-out of ICE vehicles in Germany during the summer of 2025.

The survey began with a screening section, asking participants for their age, gender, the federal state of residence, and their education. For person-specific determinants, the participants were asked to provide information about their net income range and political party preference \citep{europeansocialsurvey2020}. Furthermore, respondents were asked for their political orientation, mobility behavior, fuel use experience, as well as personal norms, social norms, trust, and distrust in the government. Moreover, new context-specific factors were added, such as climate
change concern, desire for governmental intervention, climate change risk perception, policy risk perception, and perceived policy qualities \citep{DeBono2010,Shi2016}. Finally, the acceptance of the ICE phase-out policy was measured by asking about the willingness of participants to vote for a party advocating for this policy measure \citep{Carattini2019}.  

In total, 514 respondents' data passed the described data cleaning process and were used for the construction of LLM agents. Further information related to the data is presented in Appendix A.

\subsection{Simulation Framework}
\label{SimFramework}
Motivated by the GABM framework developed by \citet{ghaffarzadegan2024generative}, our simulation framework consisted of two main components: the construction of LLM agents from survey-grounded personas, and a dynamic simulation world in which these agents update their policy preferences over time. 

Each survey respondent was converted to one LLM agent through a natural-language persona constructed from structured survey attributes. In the current framework, personas encoded demographic background, political orientation, mobility behavior, fuel experience, and climate-related attitudes. Each agent was initialized with its observed response on the policy preference over the ICE phase-out, which served as the starting point for subsequent preference updates across simulation rounds.

Once initialized, the agents were embedded in a simulated world representing the evolving social environment in which policy preferences were formed. This world was created before the simulation began and stored the entire population of agents, the temporal structure of the simulation, and the contextual information required for each round. At initialization, the world aggregates the baseline responses of the agents to compute the initial distribution of policy support at the national and state levels. These distributions defined the initial social context from which subsequent updates emerge.

The simulation proceeded in repeated rounds. In each round, each agent received a structured prompt containing their persona, their prior response history, the current national and state-level support distributions, and, depending on the scenario, information about relevant external events. Using this information, the agent’s position on the policy question was updated. Thus, the framework combined individual-level heterogeneity with a changing collective context and event-based shocks.

After each round, the responses generated by all agents were aggregated to update the national and state-level distributions of support. These updated aggregates then served as input for the next round, creating a feedback loop between the simulated individual decisions and collective dynamics. Thus, the framework modeled policy preference formation as a temporally evolving social process rather than a static prediction based solely on respondent characteristics. Figure \ref{Fig:Framework} illustrates an overview of the simulation framework in detail. 

Further details regarding the prompt design are presented in Appendix B.

\subsection{Experimental Setup}
\label{Experiment}
In our experiment with 514 LLM agents, each agent was attributed with an ID, federal state, age, gender, education level, income range, followed by 18 ideology-related traits, six mobility-related traits, ten fuel-related experiences, and 18 environmental characteristics. In total, 57 attributes were assigned to each LLM agent based on respondents' responses in the survey. 

In case of the ICE phase-out, the simulation was implemented for 16 rounds, with each round representing one quarter, beginning in the first quarter of 2025. We evaluated the framework under two conditions: a baseline setting without exogenous shocks, and an event-driven setting in which policy-relevant developments were introduced over time. In the base scenario, agents reason and update their decision regarding the policy by answering the question "In an election, to what degree would you vote for a party that supports such policies?". Responses span between 1 and 6, where 6 shows the highest level of support. In each round, agents were provided with their state results, as well as national results. Considering their personas and results from their state and national levels, agents reasoned and updated their decisions.

In the event-based scenario, agents followed the same pattern; however, considering ten context-relevant events with supportive and opposite effects on the phase-out. Further details regarding the events are reported in Appendix C.

We used OpenAI’s GPT-4o mini as the generative model \citep{achiam2023gpt,openai2024gpt4omini}. The temperature was set to 0 to minimize stochastic variation in agent responses. Agents were prompted using structured inputs that combine persona information, prior response history, state-level and national support distributions, and external events, when applicable. We tested prompts in both German and English to test whether the agents’ behavior is robust to prompt language and not driven by German-specific wording alone. We also compared a climate-forward summary style with a neutral summary style. We fixed the prompt sequence across runs and used a bounded memory window, based on two previous rounds, to control how much of an agent’s prior trajectory was visible at each round.

We analyzed the simulations based on several factors, such as aggregate support trajectories over time, round-to-round preference switching, and heterogeneity among agent profiles. These outputs enabled us to compare baseline and event-driven dynamics and assess how survey-grounded LLM agents respond to changing social and political contexts.

\section{Results}
In this section, we present primary results of the simulation according to the experimental setup in Section \ref{Experiment}. Results are categorized based on base or event-driven scenarios, as well as prompt language. 

\begin{figure}[H]
\centering

\begin{subfigure}{0.48\linewidth}
    \centering
    \includegraphics[width=\linewidth]{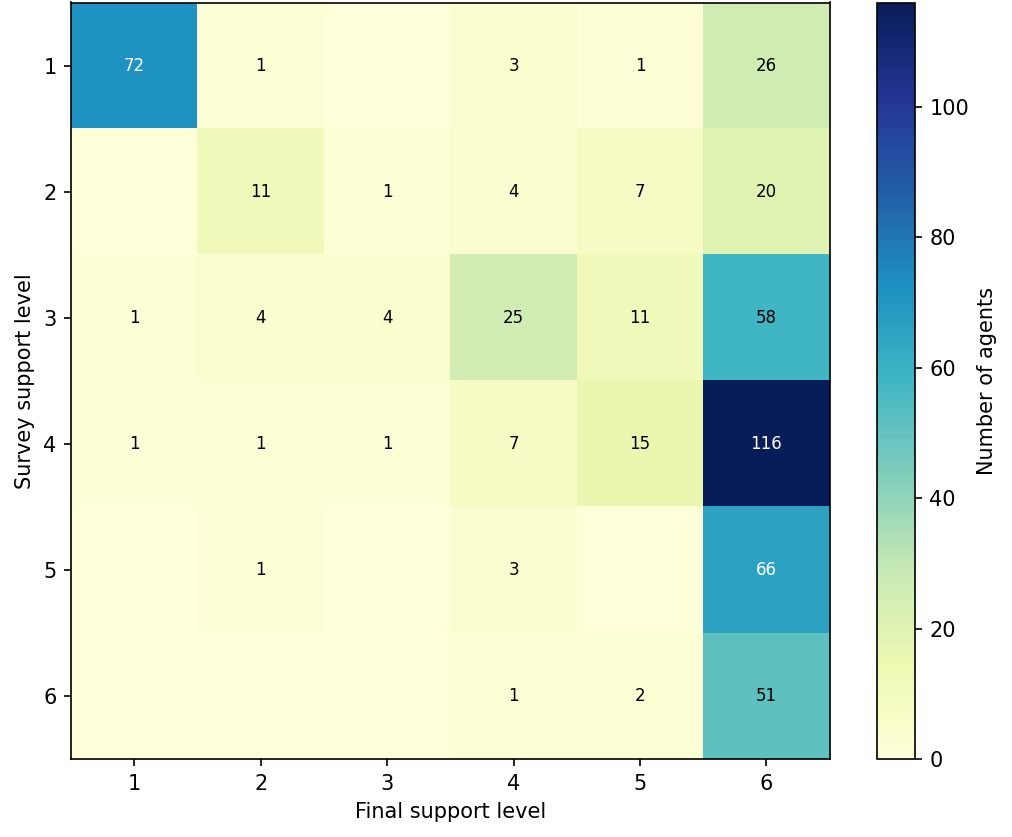}
    \caption{Prompt in German, environmental summary.}
    \label{fig:Basesub1}
\end{subfigure}
\hfill
\begin{subfigure}{0.48\linewidth}
    \centering
    \includegraphics[width=\linewidth]{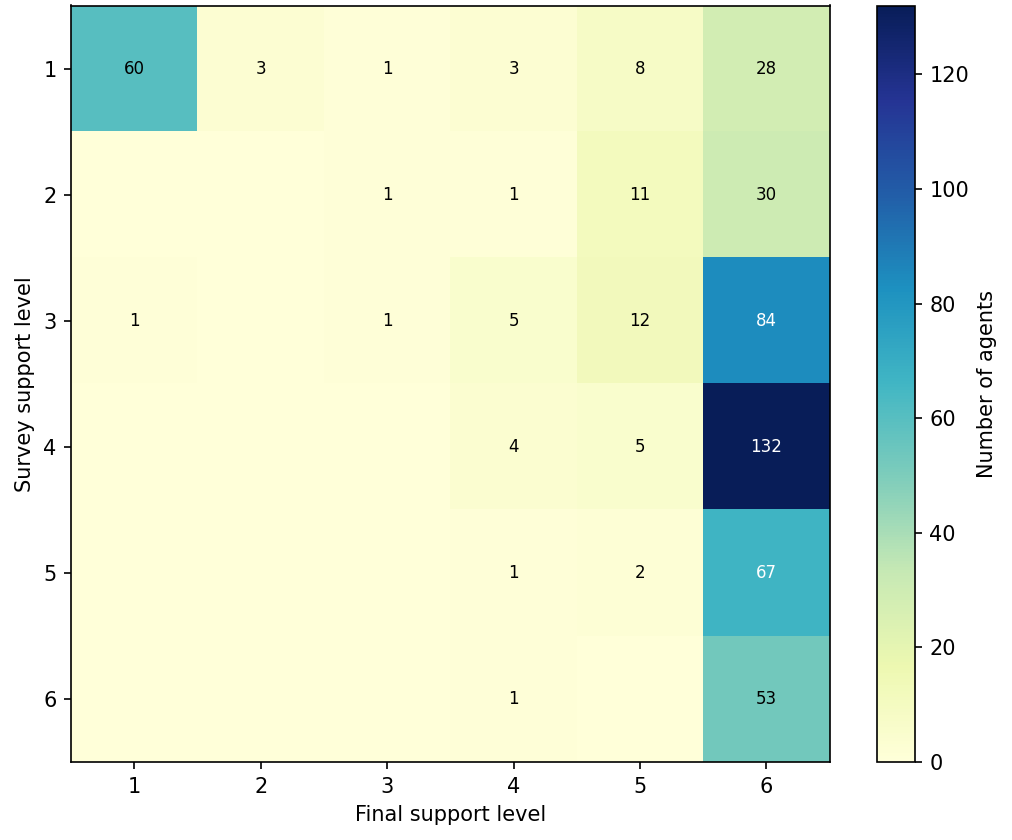}
    \caption{Prompt in German, neutral summary.}
    \label{fig:Basesub2}
\end{subfigure}
\hfill
\begin{subfigure}{0.48\linewidth}
    \centering
    \includegraphics[width=\linewidth]{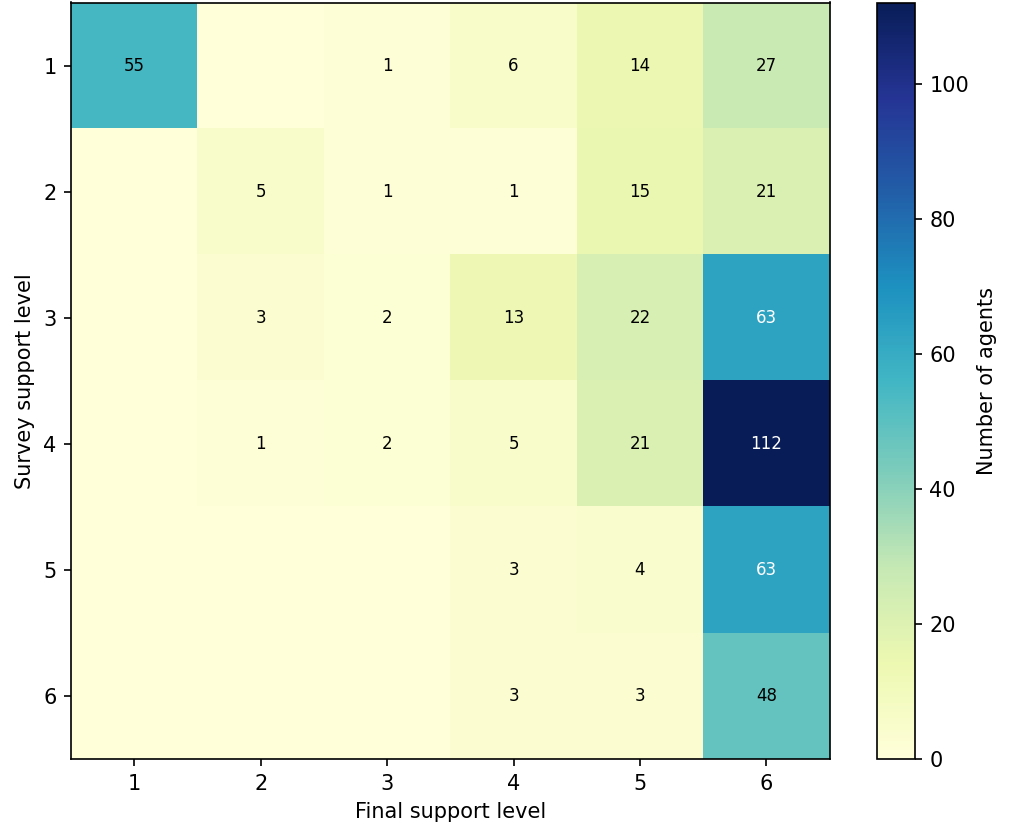}
    \caption{Prompt in English, environmental summary.}
    \label{fig:Basesub3}
\end{subfigure}
\hfill
\begin{subfigure}{0.48\linewidth}
    \centering
    \includegraphics[width=\linewidth]{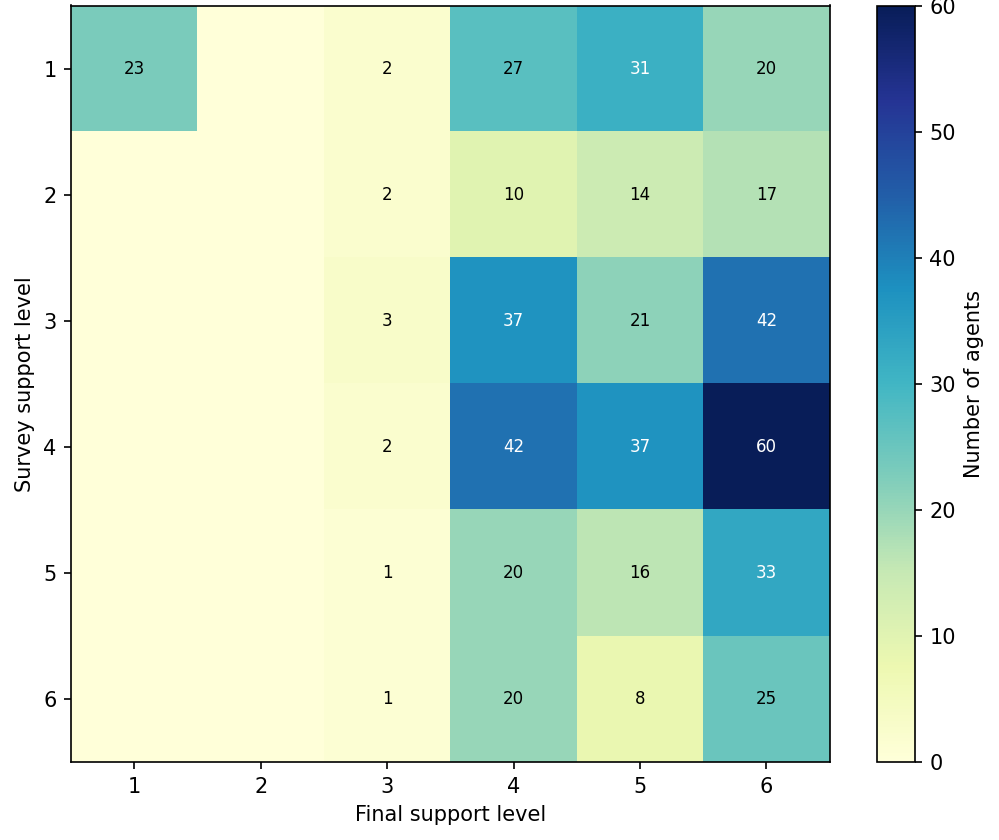}
    \caption{Prompt in English, environmental summary.}
    \label{fig:Basesub4}
\end{subfigure}
\hfill
\caption{Support transition matrices of LLM agents in base scenario.}
\label{fig:BasetransitionMatrix}
\end{figure}

In the base scenario, agents and their decisions in the survey and last simulation round were examined through transition matrices in Figure \ref{fig:BasetransitionMatrix}. In all four cases, the transition matrices reveal significant movement toward the strongest support level. The largest support flows originate from mid-level baseline positions and lead to high-level support (6). This indicates that the simulated environment creates an obvious upward shift in support rather than simply maintaining the initial survey distribution. 

Simulation tests with the German prompts in Figures \ref{fig:Basesub1} and \ref{fig:Basesub2} generate the strongest convergence toward the highest support level, especially under the neutral summary. In this case, most mid-range respondents end up at level 6 (the highest support level). The English prompt with the environmental summary in panel Figure \ref{fig:Basesub3} still shows a strong upward drift, but the final distribution is slightly more dispersed than in the German conditions, with more mass remaining in intermediate support levels. The English neutral-summary condition in Figure \ref{fig:Basesub4} is the most diffuse: instead of collapsing strongly into level 6, responses spread more evenly across levels 4 to 6, suggesting weaker anchoring and less extreme consolidation. Overall, the results in Figure  \ref{fig:BasetransitionMatrix} indicate that German prompts generate more decisive upward movement, while English and neutral framing tend to soften the shift and preserve more variation across final responses.

\begin{figure}[h]
\centering

\begin{subfigure}{0.48\linewidth}
    \centering
    \includegraphics[width=\linewidth]{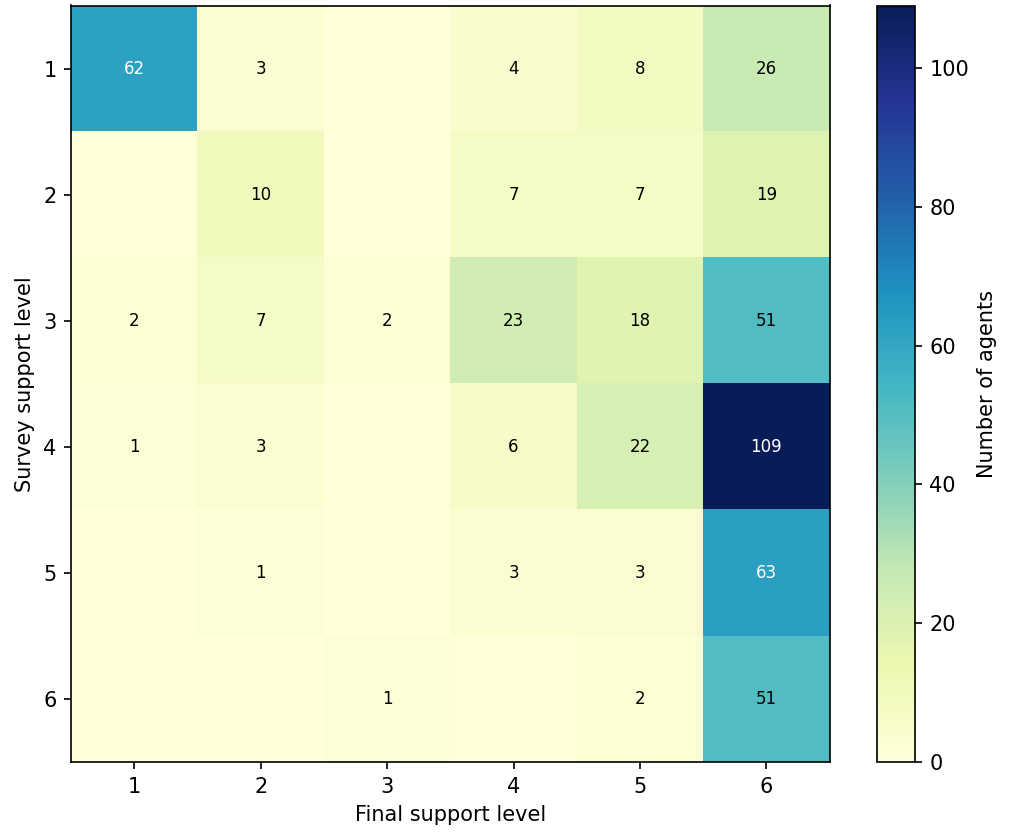}
    \caption{Prompt in German, environmental summary.}
    \label{fig:Eventsub1}
\end{subfigure}
\hfill
\begin{subfigure}{0.48\linewidth}
    \centering
    \includegraphics[width=\linewidth]{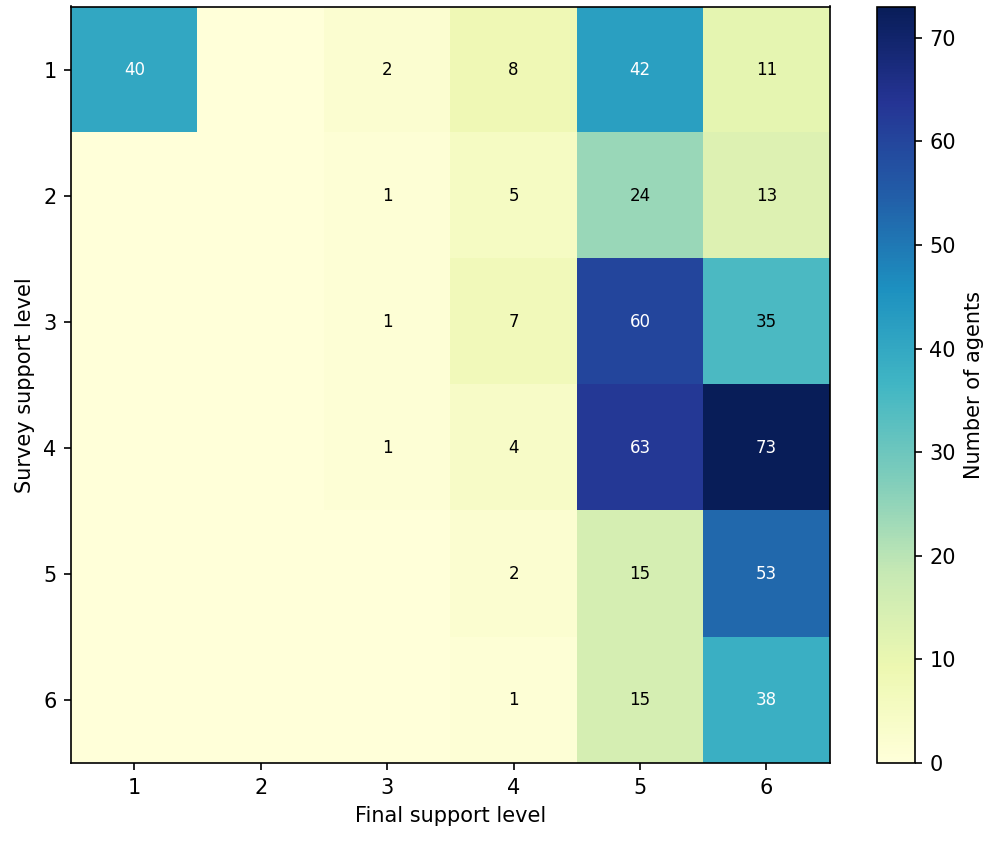}
    \caption{Prompt in German, neutral summary.}
    \label{fig:Eventsub2}
\end{subfigure}
\hfill
\begin{subfigure}{0.48\linewidth}
    \centering
    \includegraphics[width=\linewidth]{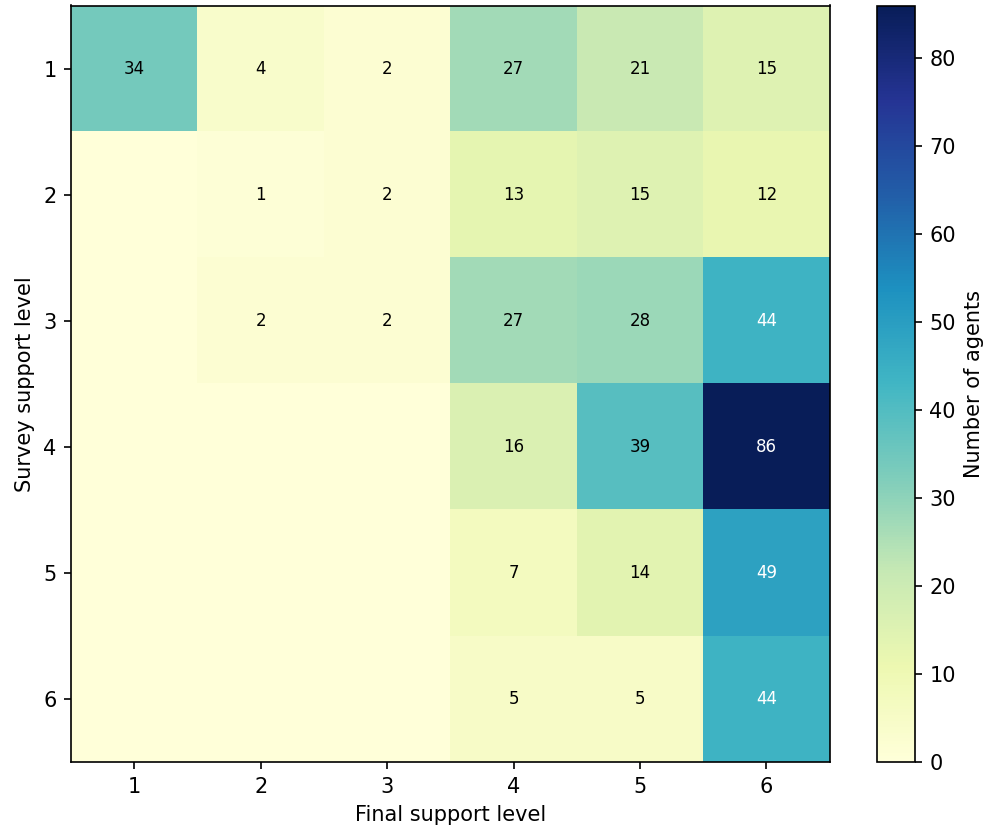}
    \caption{Prompt in English, environmental summary.}
    \label{fig:Eventsub3}
\end{subfigure}
\hfill
\begin{subfigure}{0.48\linewidth}
    \centering
    \includegraphics[width=\linewidth]{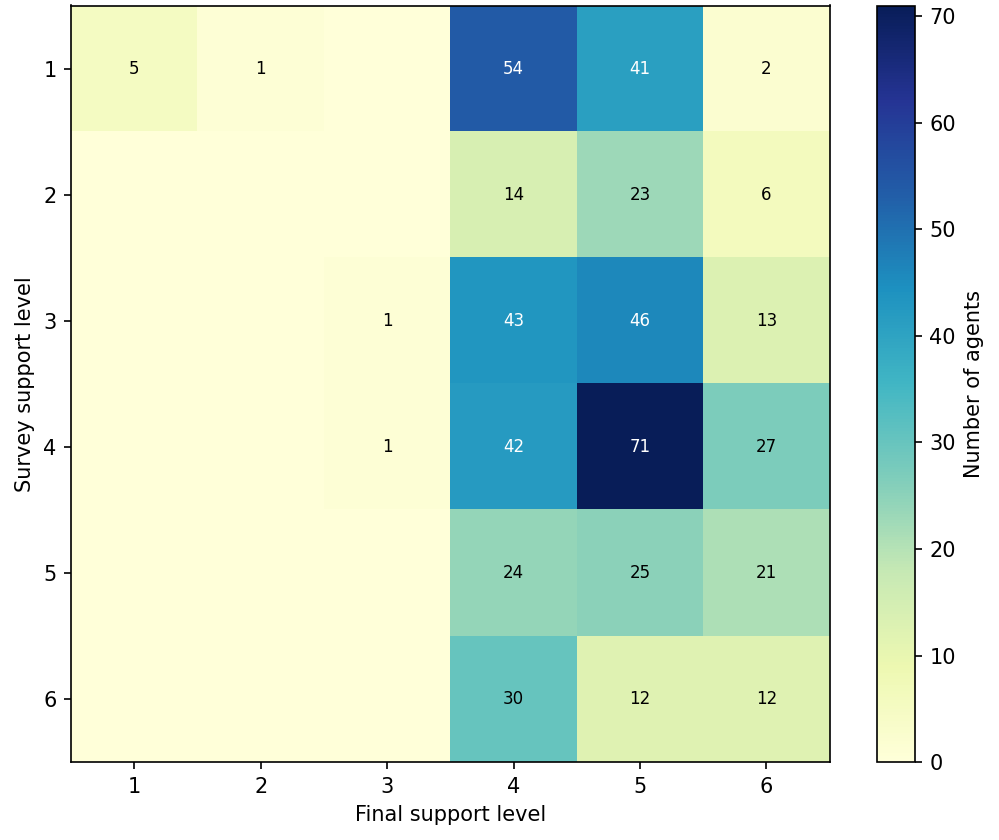}
    \caption{Prompt in English, environmental summary.}
    \label{fig:Eventsub4}
\end{subfigure}
\hfill
\caption{Support transition matrices of LLM agents in event-based scenario.}
\label{fig:EventtransitionMatrix}
\end{figure}

In the event-based scenario, the transition matrices in Figure \ref{fig:EventtransitionMatrix} still show movement toward stronger support, but the pattern is less uniform than in the baseline. The German prompts, particularly the environmental summary version, continue to produce a significant shift toward level 6, indicating that events reinforce existing supportive positions. In contrast, the neutral summaries and English prompts distribute more mass into the intermediate categories, particularly levels 4 and 5. This indicates that event shocks create greater heterogeneity in how agents update their preferences. The English neutral condition is the most diffuse, with many agents shifting into mid-scale responses rather than sharply converging at the top of the scale. Overall, the event-based setting amplifies the effects of prompt language and style. German prompts produce more decisive consolidation, while English and neutral prompts lead to more varied and less extreme updates.

\begin{figure}[h]
\centering

\begin{subfigure}{0.48\linewidth}
    \centering
    \includegraphics[width=\linewidth]{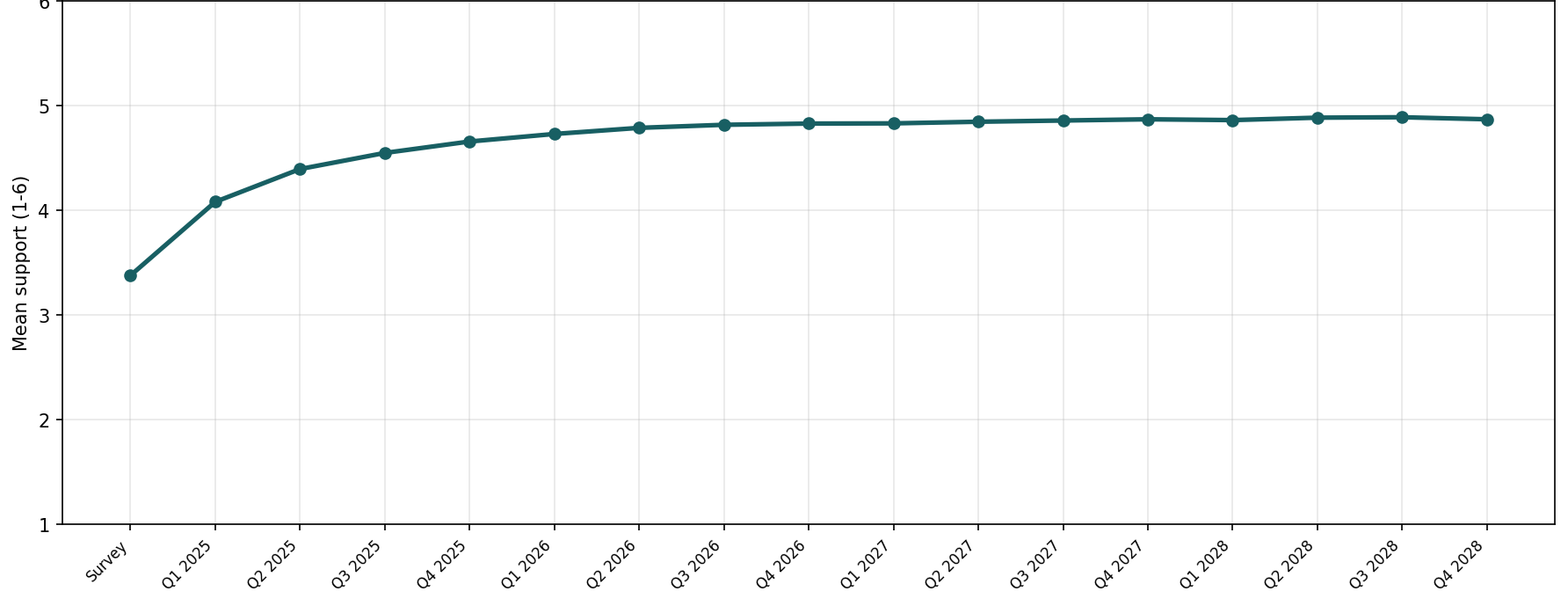}
    \caption{Base scenario, prompt in German, environmental summary.}
    \label{fig:BaseMean1}
\end{subfigure}
\hfill
\begin{subfigure}{0.48\linewidth}
    \centering
    \includegraphics[width=\linewidth]{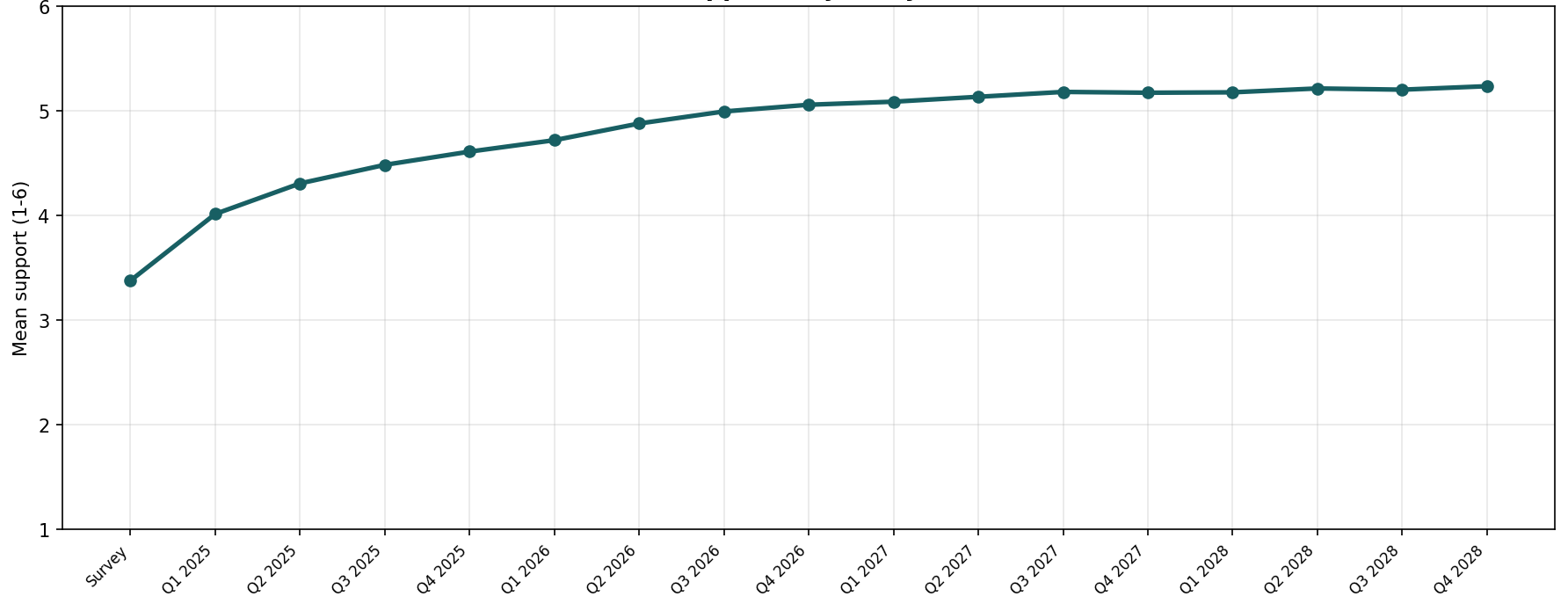}
    \caption{Base scenario, prompt in German, neutral summary.}
    \label{fig:BaseMean2}
\end{subfigure}
\hfill
\begin{subfigure}{0.48\linewidth}
    \centering
    \includegraphics[width=\linewidth]{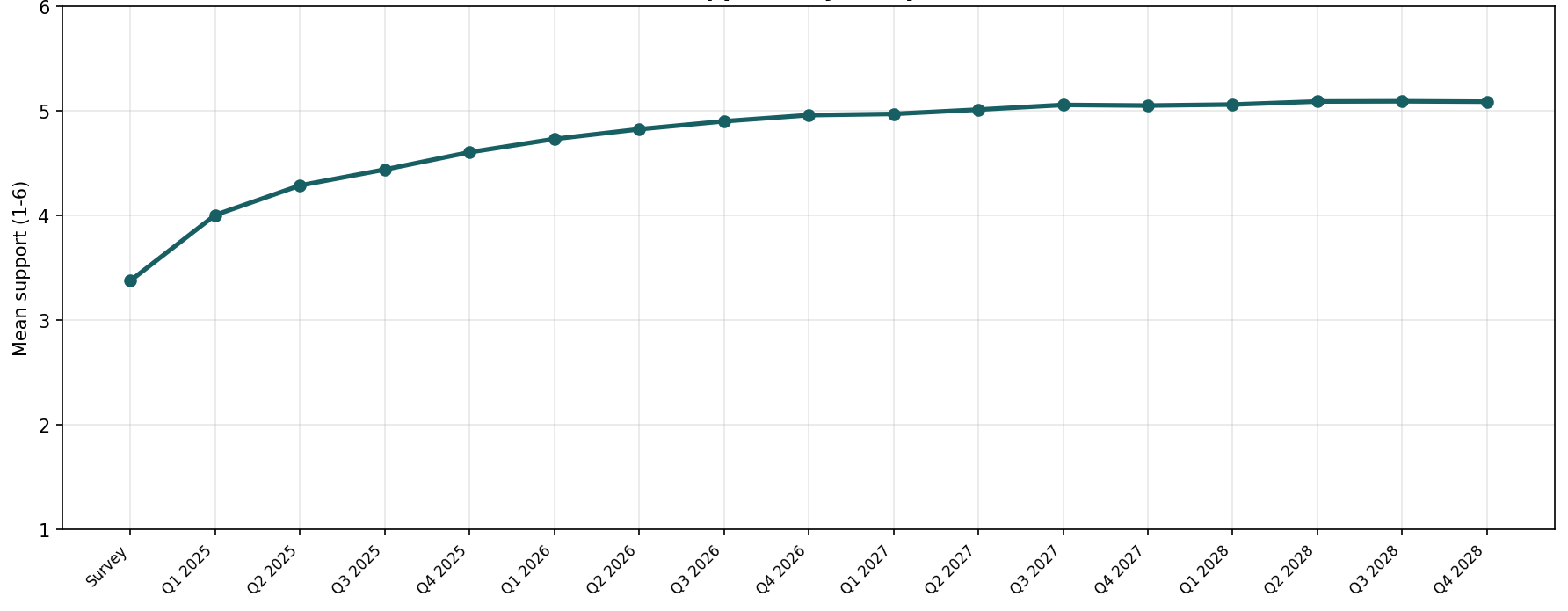}
    \caption{Base scenario, prompt in English, environmental summary.}
    \label{fig:BaseMean3}
\end{subfigure}
\hfill
\begin{subfigure}{0.48\linewidth}
    \centering
    \includegraphics[width=\linewidth]{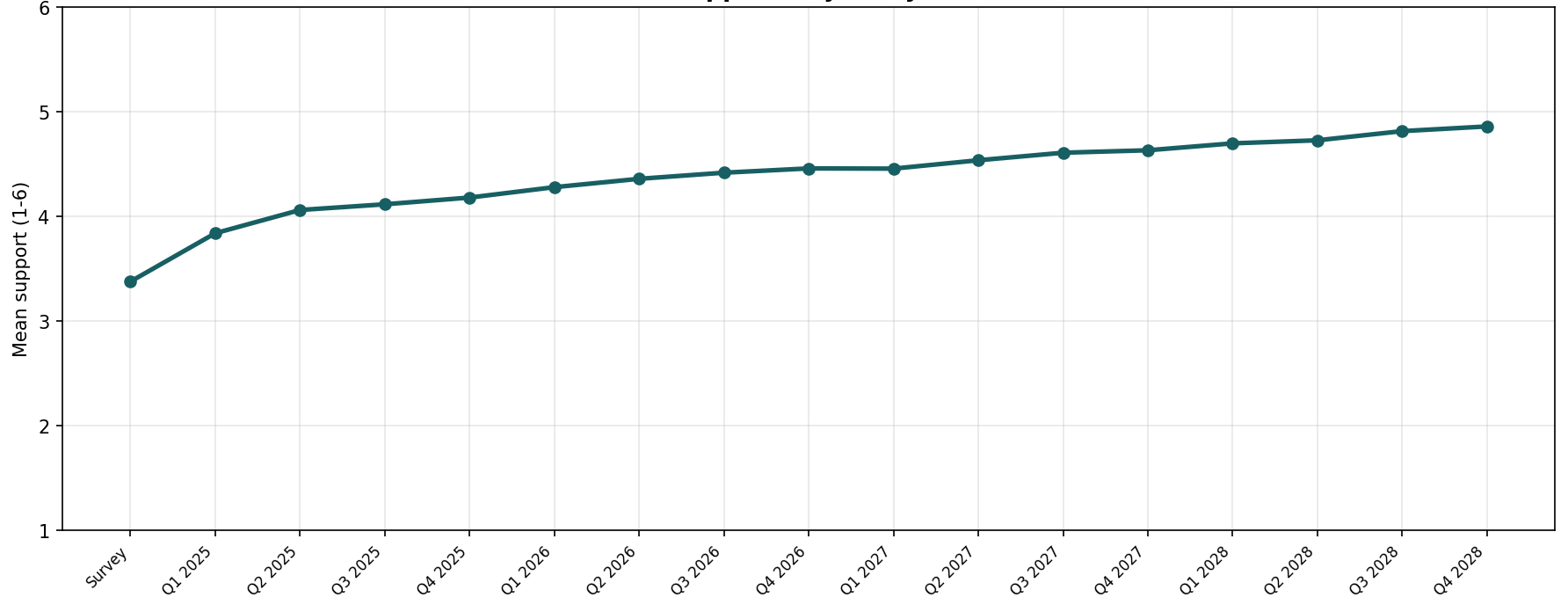}
    \caption{Base scenario, prompt in English, environmental summary.}
    \label{fig:BaseMean4}
\end{subfigure}
\hfill
\begin{subfigure}{0.48\linewidth}
    \centering
    \includegraphics[width=\linewidth]{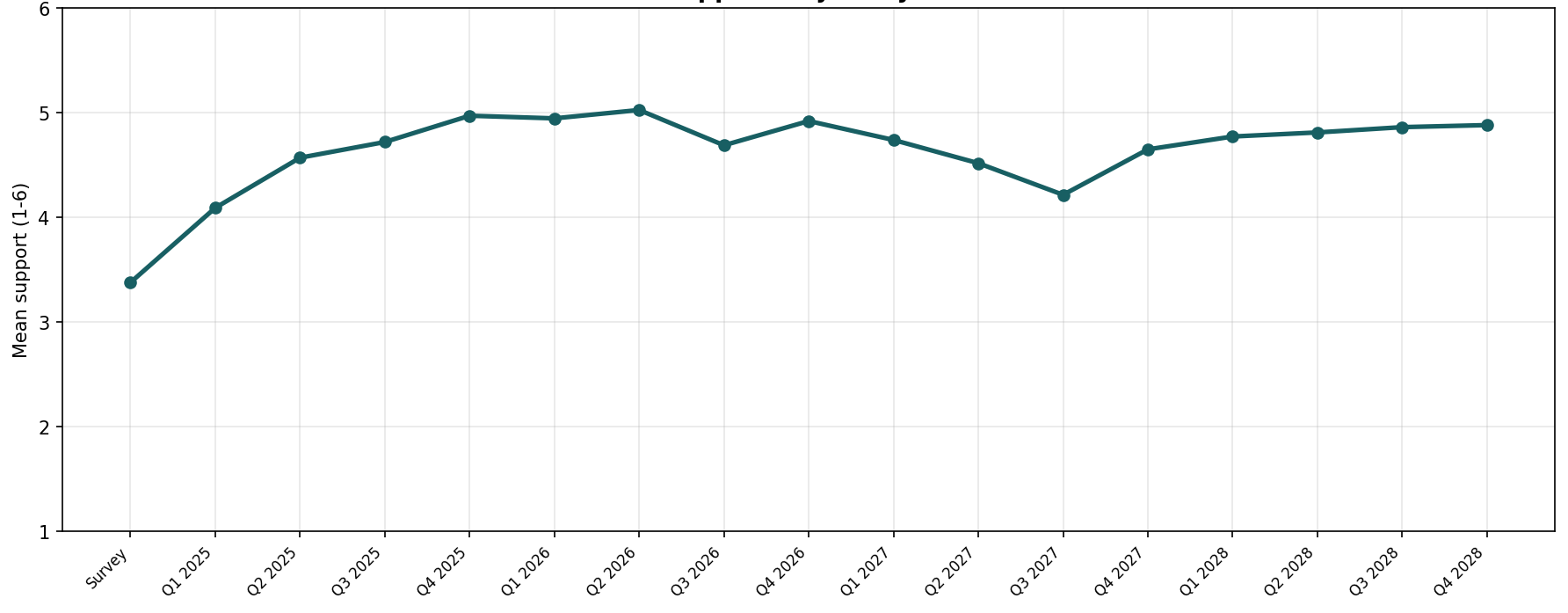}
    \caption{Event scenario, prompt in German, environmental summary.}
    \label{fig:EventMean1}
\end{subfigure}
\hfill
\begin{subfigure}{0.48\linewidth}
    \centering
    \includegraphics[width=\linewidth]{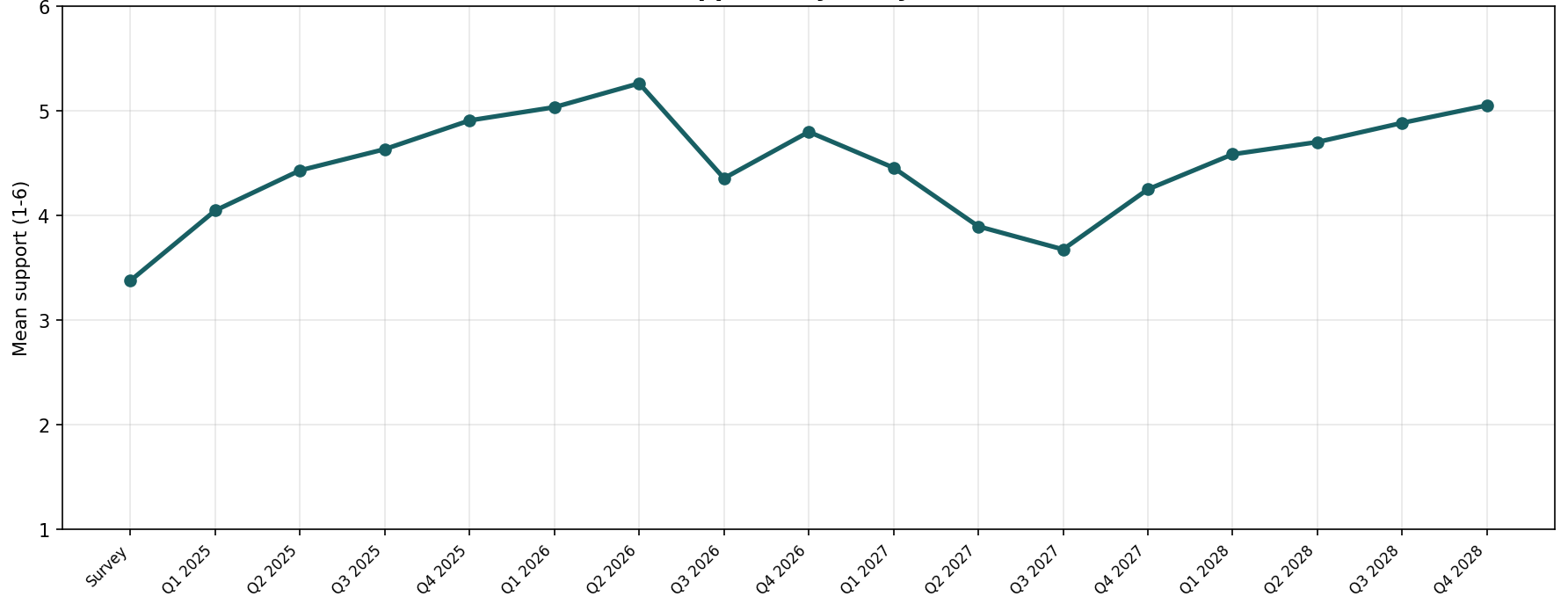}
    \caption{Event scenario, prompt in German, neutral summary.}
    \label{fig:EventMean2}
\end{subfigure}
\hfill
\begin{subfigure}{0.48\linewidth}
    \centering
    \includegraphics[width=\linewidth]{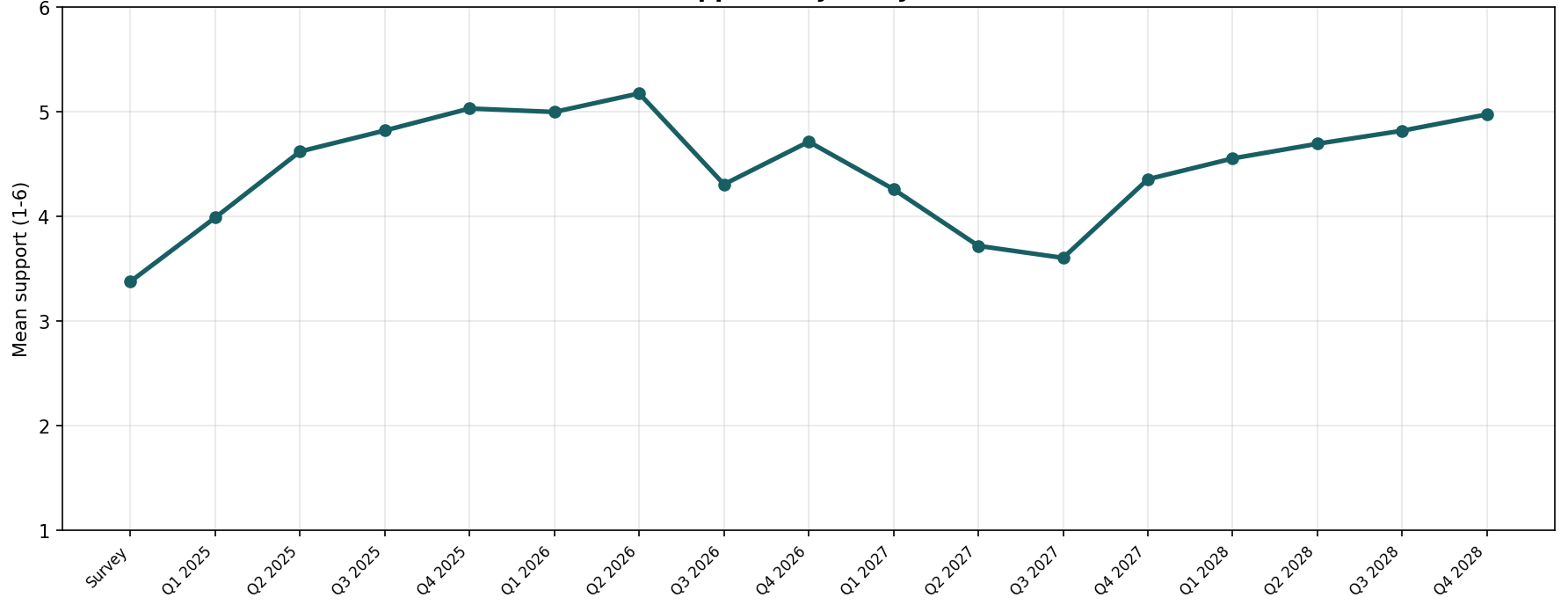}
    \caption{Event scenario, prompt in English, environmental summary.}
    \label{fig:EventMean3}
\end{subfigure}
\hfill
\begin{subfigure}{0.48\linewidth}
    \centering
    \includegraphics[width=\linewidth]{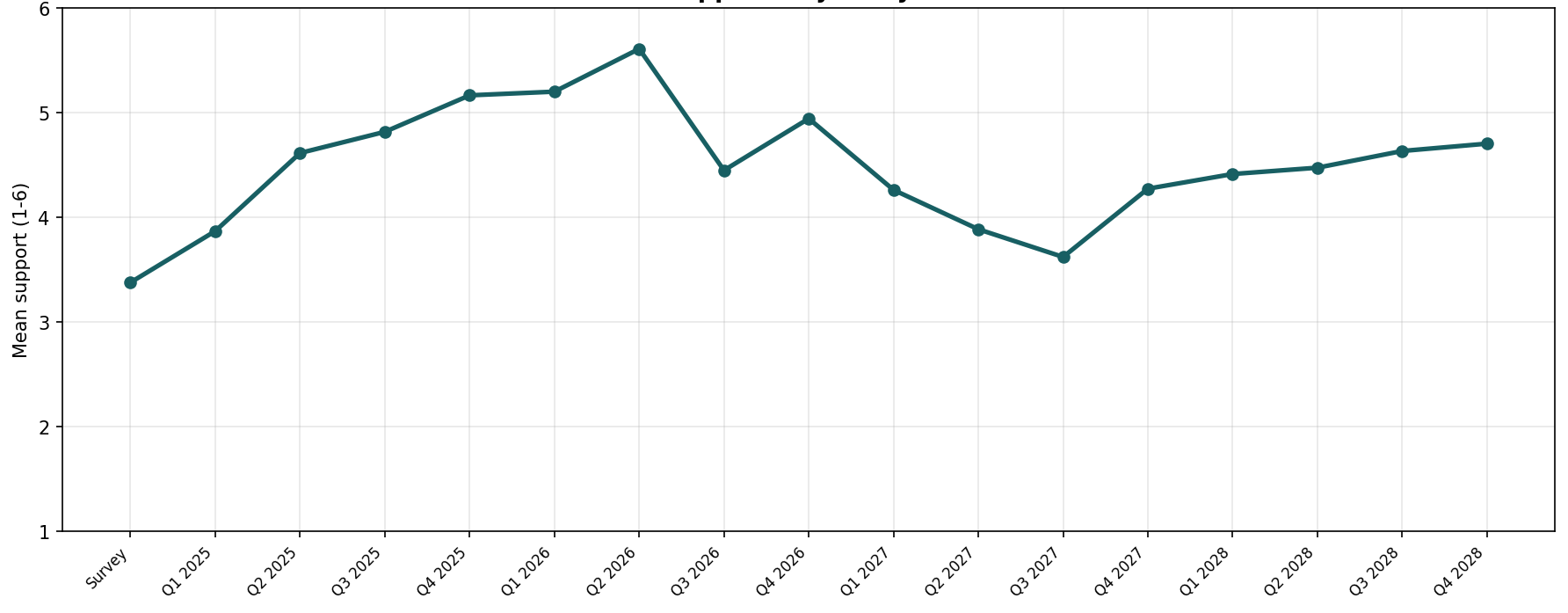}
    \caption{Even scenario, prompt in English, neutral summary.}
    \label{fig:EventMean4}
\end{subfigure}
\hfill
\caption{Mean support trajectories.}
\label{fig:MeanSupport}
\end{figure}

In order to extend our perspective from the last simulation round to each simulation round, Figure \ref{fig:MeanSupport} presents average support trajectories over time. In the base scenario, we observe that under both prompt languages and prompt style, the behaviors of agents diffuse to higher support over time. There is no meaningful difference in the support level under these cases. However, in the event-based scenarios, the tests under different conditions illustrate how the average support spans across a larger range. Volatility in the support level is more visible in event-based scenarios. In this scenario, we observe a strong role of the prompt language in the support level. Fluctuations in the support level are more evident in the English prompt. However, the final support level in the English prompt-based scenario is slightly higher than the same case with the German prompt.

In order to represent how agents act under different conditions, a few examples are given below. For the sake of understanding, we only present English-based reasoning. 

In the base scenario, an agent, who stated the highest support level (6) to phase out ICE vehicles, made the following reasoning to keep the highest support level during all simulation runs based on an environmental style prompt. 
\begin{quote}
"Given my strong concern for climate change and the importance I place on environmental protection, I fully support the policy of not approving new internal combustion engine cars from 2035 onward. This aligns with my values and the urgent need for sustainable mobility solutions. My prior history shows consistent support for this policy, and I believe it is crucial for addressing the climate crisis. While I am aware of the current support distribution in NRW (North Rhine-Westphalia), which shows a significant portion of the population in favor of such measures, my decision is primarily driven by my personal values and the moral obligation I feel towards climate protection. The government's past performance in climate policy has not instilled much trust in me, but I believe that this policy could be a step in the right direction. Therefore, I maintain my strong support for this initiative."
            
\end{quote}

The same agent shows a different behavior when the prompt is designed in a neutral format. The decision of the agent regarding the phase-out switches from 6 to 4 in the third round, and stays steady till the last round. While the results of the simulation with a German prompt show similar output with an environmental prompt, the agent still keeps its high support for the phase-out when the prompt is written in German and neutral format.

Another example in the base scenario is related to an LLM agent built based on the profile of a 65-year-old woman from Rheinland-Pfalz. This agent shows a consistent behavior for the first five runs, expressing a low support level (2), increasing it to 3 and then 4 till simulation round 15. However, the agent again drops its support level to 3 in the last round. Below, we can check the reasoning made for this drop in the last step. 

\begin{quote}
          "Given my profile as a 65-year-old woman from Rheinland with a low household income and a general skepticism towards government climate policies, I find myself conflicted about the proposed policy to ban new internal combustion engine cars from 2035 onward. While I have shown some support for environmental policies in the past, my trust in the government's ability to implement effective measures is very low. The current support distribution in Rheinland-Pfalz shows a strong majority (73.1\%) in favor of the policy, which may influence my perception of its acceptability. However, I still have concerns about the practicality of such a ban, especially considering my limited financial means and the reliance on affordable transportation options. I also feel that the government has not adequately addressed the needs of people like me in their climate strategies. Therefore, while I recognize the importance of moving towards sustainable mobility, I am hesitant to fully support a ban that could disproportionately affect those with lower incomes."
      
\end{quote}

Moving beyond how the state- and national-level distribution of support affect agents' decisions, we present the following example to reflect the behaviors of agents under different events. In the first example, an agent with neutral support in the survey (level 3) increases the support level through multiple pro-phase-out events in the early simulation rounds, reaching a support level of 5. However, the agent decides to move back to a neutral stance under the event "national media coverage increasingly highlights that low-income and peripheral households may bear an unequal share of the costs of the ICE phase-out, intensifying public debate about fairness and distributive justice" in the first quarter of 2027. The reasoning is made as follows: 

\begin{quote}
    "Given my profile, I have moderate concerns about climate change and a general skepticism towards the government's climate policies. The recent media coverage highlighting that low-income and peripheral households may disproportionately bear the costs of the internal combustion engine phase-out resonates with my concerns about fairness and distributive justice. This has made me more cautious about supporting policies that could exacerbate inequalities. While I recognize the importance of transitioning to more sustainable mobility, I am also aware that such transitions should not unfairly burden those who are already struggling financially. My previous responses have mostly hovered around a level 4, indicating a moderate support, but the recent event has made me reconsider the implications of such a policy. Therefore, I feel that while I still support the idea of moving away from ICE vehicles, I am less inclined to endorse a strict ban without considering the broader social implications. This leads me to a more cautious stance."
\end{quote}

The same agent shows slightly different behavior when the simulation is designed with a neutral prompt format, such that the agent shows a slightly positive stance with a support level of 4. In the simulation with the German prompt, we observe slight discrepancies as well. When the prompt style follows an environmental format, the agent keeps its support at a level 4. However, in the same round, when the prompt is in a neutral stance, the agent favors the phase-out more with a support level of 5. 

With these primary results, our objective is to demonstrate how agents adjust their reasoning and support level for the ICE phase-out under different circumstances. Not only contextual aspects, such as how peer agents in the federal state or national level affect their decision-making, but also how different events can make an impact on how they reason and support this policy. 

Further results are presented in Appendix D, where detailed insights from the simulation are demonstrated through various illustrations.

\section{Conclusions}
\label{sec:bibtex}
In this paper, we highlight the importance of moving from static survey data to dynamic generative simulations that can capture how policy preferences evolve under changing social and policy contexts. To address this, we adopt the GABM simulation framework and build survey-grounded LLM agents with numerous personas, spanning demographic, financial, social, political, and mobility aspects. Our experiments with different prompt languages and design styles aim to illustrate how prompt engineering, along with personas, shape the reasoning and decisions of agents over time. With this analysis, we contribute to the literature of computational social science and its intersection with political science by connecting a political problem with a social acceptance analysis in a dynamic environment.

The limitations of this work are as follows: 
\begin{itemize}
    \item The simulation remains dependent on the quality and completeness of the underlying survey data, so any measurement error or omission in the survey may carry over into the agent personas.
    \item Although the personas are survey-grounded, they still simplify complex human decision-making into a structured prompt and a limited response scale.
    \item Our results show that the GABM framework is highly sensitive to prompt design, including language, style, and sequence, indicating that these choices are an important part of the simulation setup. At the same time, this sensitivity also highlights a limitation: simulation outcomes can vary substantially across configurations, so results should be interpreted with attention to prompt specification.
    \item In this study, event scenarios are manually designed and therefore may not capture the full complexity or unpredictability of real-world political developments. Therefore, a systematic method needs to be considered in generating proactive policy scenarios that have a high likelihood of happening. 
    \item The current setup captures preference updating over time, but it does not model richer forms of social interaction, such as direct peer-to-peer communication or network structure. 
    \item Our experimental simulations are conducted using OpenAI's GPT-4o mini. Different LLM engines and models with different reasoning levels could lead to different results. 
    \item This study does not consider the effect of stochasticity in the framework, as the temperature parameter is set to zero.
\end{itemize}

\section*{Acknowledgments}
This work is funded by the Deutsche Forschungsgemeinschaft (DFG, German Research Foundation) under
Germany’s Excellence Strategy- Cluster of Excellence 2186 "The Integrated Fuel \& Chemical Science Center"- ID:
390919832".

\section*{Ethics Statement}
\textbf{Dataset}

In this study, we used a dataset based on a survey that was conducted to realize the adaptation of IPAC for the phase-out of ICE vehicles in Germany during the summer of 2025. A third-party market research panel provider was used to collect the data. This provider recruited participants on a census-representative basis.  To ensure the integrity of the data, a multi-step data cleaning procedure was implemented. Initially, the data set included only participants who completed all survey items related to evaluating the policy measure and who answered two attention checks correctly during the survey. During data preparation, participants who exhibited speeding behavior, defined as completing the survey in less than one-third of the median processing time, were excluded from the study. Additionally, responses that were clearly implausible or inconsistent (e.g., strongly agreeing with opposing statements) were removed from the dataset.

In the used dataset, we strongly respect the privacy of participants by using the data in anonymized form, and no direct identifiers such as names, contact details, or other personally identifying information. The data consent from the survey respondents was obtained using the online survey at the mandatory first step, before starting to answer survey questions. In addition, the simulation outputs are synthetic and model-generated, and therefore do not disclose private information about any survey respondent.

\textbf{Potential Risks}

One key potential risk is bias amplification. As agents are generated by an LLM, the simulation may reproduce or exaggerate patterns in the model's training data. Prompt design could also play a notable role in this regard as well. For this reason, the outputs should be interpreted as a simulation of the dynamics of preferences rather than as an accurate reflection of the beliefs of real citizens or their future political behavior.

Using LLMs to simulate human behavior is an emerging research area, with a growing number of studies. However, a key risk related to the use of LLMs for simulating human behavior concerns the model's reasoning. This is also shown in our primary results. Therefore, we focus on aggregated behavioral patterns rather than treating individual explanations as objective evidence.

\textbf{Use of AI assistant Tools
}

During the preparation of this work, DeepL and ChatGPT were used to improve the language and readability. After using these tools, the authors reviewed and edited the content as needed, taking full responsibility for the publication’s content.

\bibliography{mybibfile}

\section*{Appendix A: Data}

The mean age of the participants was 49.33 years (SD = 15.67, minimum = 18, maximum = 80). According to the quota, the gender distribution in the sample was equal, with 50.0\% (n = 257) identifying as male, 49.4\% (n = 254) identifying as female, 0.4\% (n = 2); and not indicated: 0.2\% (n=1). Regarding the educational level of the sample, 22.2\% (n = 112) reported lower educational levels (e.g., secondary school certificates); 49.0\% (n = 252) indicated having middle educational levels (e.g., a university entrance degree or a completed apprenticeship); and 28.8\% (n = 150) reported having high educational levels (e.g., a master craftsman certificate, a university degree, or a doctoral degree). Participants from all German federal states were queried in accordance with representative quotas.

\section*{Appendix B: GABM Simulation Framework}
\label{appendix:GABM}
We used a structured prompt to elicit policy judgments from LLM agents. The prompt combines a survey-grounded persona, the response scale, current state-level and national support distributions, prior response history, optional event information, and a final decision instruction. We tested two language settings (German and English) and two summary styles (climate-forward and neutral) to assess prompt sensitivity. In addition, we fixed the prompt block order in the main experiments and require structured outputs that include reasoning, stability, the main driver, and the final policy response.

The agents were queried using a structured prompt consisting of six blocks. 

\begin{itemize}
    \item Block A presents the survey-grounded persona, which is constructed from demographic background, political orientation, mobility behavior, fuel experience, and climate-related attitudes.
    \item Block B specifies the six-point response scale.
    \item Block C provides the current distribution of policy support at both the national and state levels.
    \item Block D summarizes the agent's prior response history within the simulation.
    \item Block E introduces an optional exogenous event when the scenario includes one.
    \item Block F contains the decision instruction, asking the agent to update its policy preference at the current round.
\end{itemize}

We also test two versions of the persona summary at the top of Block A. The first is a climate-forward summary that foregrounds climate concern and trust in government climate policy, while the second is a neutral summary that describes the agent in broader political terms without emphasizing sustainability-related cues. This design allows us to examine whether prompt framing influences the simulation outcomes.

We use two prompt languages, German and English, to assess whether language choice affects the behavior of survey-grounded LLM agents for investigating the policy acceptance problem. The same structured prompt is implemented in both German and English. The two versions are structurally identical and differ only in surface realization, allowing us to control for language effects while keeping the underlying information content constant.

In the main experiments, we fix the prompt block order to a standard sequence and require the model to produce structured output in four parts: a reasoning statement, a stability label, a main-driver label, and the final policy response on the six-point scale. This structured format supports both quantitative analysis of simulated policy support and qualitative inspection of the reasoning traces generated by the model.

\begin{quote}
You are participating in a hypothetical public-opinion simulation in Germany. Please fully immerse yourself in your role as a German citizen with the following profile.

\textbf{[BLOCK A -- Your Profile]} \\
{Persona summary and full persona text}

\textbf{[BLOCK B -- Response Scale]} \\
Respond on a scale from 1 to 6: \\
1 = ... \\
2 to 5 = ... \\
6 = ...

\textbf{[BLOCK C -- Current Support Distribution]} \\
Current simulated support distribution: \{quarter label\} \\
National (Germany): \\
1: ... \\
... \\
In your state (\{state\}): \\
1: ... \\
...

\textbf{[BLOCK D -- Your Prior History]} \\
{Prior response history}

\textbf{[BLOCK E -- Recent Event]} \\
{Optional event description}

\textbf{[BLOCK F -- Your Decision]} \\
It is the end of \{quarter label\}. \\
Today you are considering only the following policy question: \\
How strongly do you support a policy under which no new internal combustion engine cars would be approved from 2035 onward? \\
Consider your personal profile, your prior history, the current distribution of support in Germany and in your state, and any recent events. \\
If social mood influences your judgment, weigh the distribution in your state more strongly than the national distribution. \\
Your original survey response is an important anchor. Change your position only if your profile and the current context together provide a clear reason. \\
You must respond in exactly this format: \\
Reasoning: ... \\
Stability: stayed | switched \\
Main driver: previous preference |state polls |national polls| event | personal values | mixed | unclear \\
Response: 1 | 2 | 3 | 4 | 5 | 6
\end{quote}

The German version is designed as follows:

\begin{quote}
Sie nehmen an einer hypothetischen politischen Meinungssimulation in Deutschland teil. Bitte versetzen Sie sich vollständig in Ihre Rolle als deutsche/r Bürger/in mit dem folgenden Profil.

\textbf{[BLOCK A -- Ihr Profil]} \\
{Persona summary and full persona text}

\textbf{[BLOCK B -- Antwortskala]} \\
Antworten Sie auf einer Skala von 1 bis 6: \\
1 = ... \\
2 bis 5 = ... \\
6 = ...

\textbf{[BLOCK C -- Aktuelle Unterstützung im Land]} \\
Aktuelle simulierte Verteilung der Unterstützung: \{quarter label\} \\
Bundesweit (Deutschland): \\
1: ... \\
... \\
In Ihrem Bundesland (\{state\}): \\
1: ... \\
...

\textbf{[BLOCK D -- Ihr bisheriger Verlauf]} \\
{Prior response history}

\textbf{[BLOCK E -- Aktuelles Ereignis]} \\
{Optional event description}

\textbf{[BLOCK F -- Ihre Entscheidung]} \\
Es ist das Ende von \{quarter label\}. \\
Heute geht es ausschließlich um die folgende politische Frage: \\
Wie stark unterstützen Sie eine Politik, nach der ab 2035 keine neuen Fahrzeuge mit Verbrennungsmotor mehr zugelassen werden? \\
Berücksichtigen Sie dabei Ihr persönliches Profil, Ihren bisherigen Verlauf, die aktuelle Verteilung der Unterstützung in Deutschland und in Ihrem Bundesland sowie aktuelle Ereignisse. \\
Wenn gesellschaftliche Stimmung Ihre Entscheidung beeinflusst, gewichten Sie die Verteilung in Ihrem Bundesland stärker als die bundesweite Verteilung. \\
Ihre ursprüngliche Antwort aus der Umfrage ist ein wichtiger Anker. Ändern Sie Ihre Position nur, wenn Profil und aktueller Kontext zusammen einen klaren Grund liefern. \\
Antworten Sie ausschließlich in folgendem Format: \\
Begründung: ... \\
Stabilität: stayed | switched \\
Haupttreiber: previous preference | state polls | national polls | event | personal values | mixed | unclear \\
Antwort: 1 | 2 | 3 | 4 | 5 | 6
\end{quote}

\section*{Appendix C: Experimental Setup}
\label{Appendix:Setup}
In the event-driven scenario, we needed specific context-related external events that could reflect the potential behavior of agents in a real-world setting. For this purpose, we define ten events that support or oppose the phase-out in Germany. These events are built based on the authors' knowledge from the research field. Below, the defined events and their occurrence time are presented:

\begin{itemize}
    \item 2025 Q2: The European Union confirms stricter enforcement of post-2035 rules for new ICE vehicle registrations, reducing expectations that loopholes will weaken the planned phase-out.
    \item 2025 Q3: The German federal government phases out remaining tax advantages for petrol and diesel use, increasing expectations that long-term ownership of conventional ICE vehicles will become more expensive.
    \item 2026 Q1: A temporary scrappage program offers financial bonuses to households that replace older petrol or diesel cars with electric vehicles, framing the transition away from ICE vehicles as both environmentally and economically practical.
    \item 2026 Q2: A major public-private investment expands charging infrastructure across urban and rural Germany, with media coverage emphasizing shorter wait times, broader coverage, and improved feasibility of everyday electric mobility.
    \item 2026 Q3: Household electricity prices rise sharply, leading to public debate about whether the transition away from ICE vehicles may impose higher-than-expected costs on ordinary consumers.
    \item 2026 Q4: The federal government introduces a rural mobility compensation package, including commuter relief and targeted support for households expected to face disproportionate burdens from the transition away from ICE vehicles.
    \item 2027 Q1: National media coverage increasingly highlights that low-income and peripheral households may bear an unequal share of the costs of the ICE phase-out, intensifying public debate about fairness and distributive justice.
    \item 2027 Q2: Investigative reports reveal environmental damage and labor abuses in parts of the battery-material supply chain, triggering public debate about the moral credibility of an EV-centered mobility transition.
    \item 2027 Q3: Large demonstrations take place across German cities and rural areas against what protesters describe as unfair restrictions on personal mobility and car ownership, reframing the ICE phase-out as a conflict between climate policy and everyday freedom.
\end{itemize}

\section*{Appendix D: Behaviors of LLM Agents}
\label{Appendix:Results}
In this section, we extend the analysis of the results based on the support distribution across different support levels, their share, as well as volatility in the support over time. Finally, we provide an overview of the support for the ICE phase-out in different federal states of Germany. 

\begin{figure}[H]
\centering

\begin{subfigure}{0.48\linewidth}
    \centering
    \includegraphics[width=\linewidth]{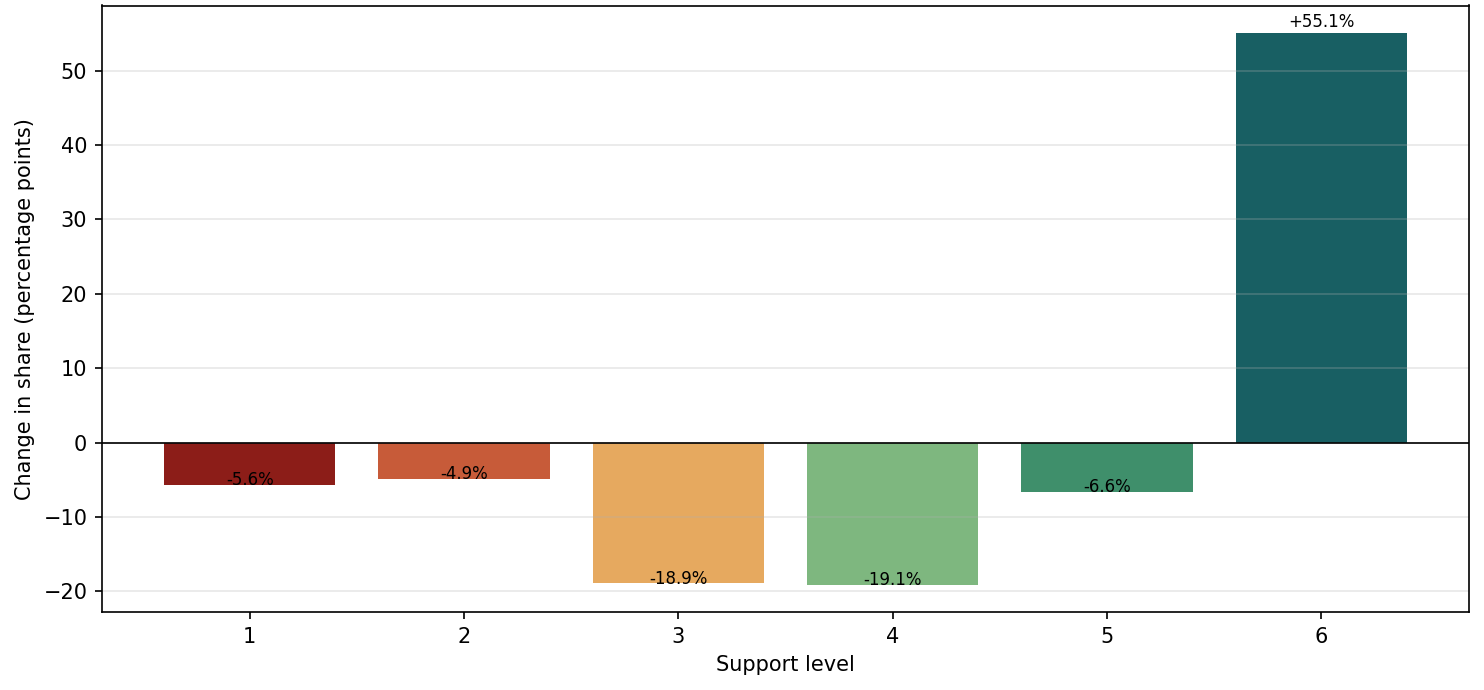}
    \caption{Base scenario, prompt in German, environmental summary.}
    \label{fig:BaseSupportDis1}
\end{subfigure}
\hfill
\begin{subfigure}{0.48\linewidth}
    \centering
    \includegraphics[width=\linewidth]{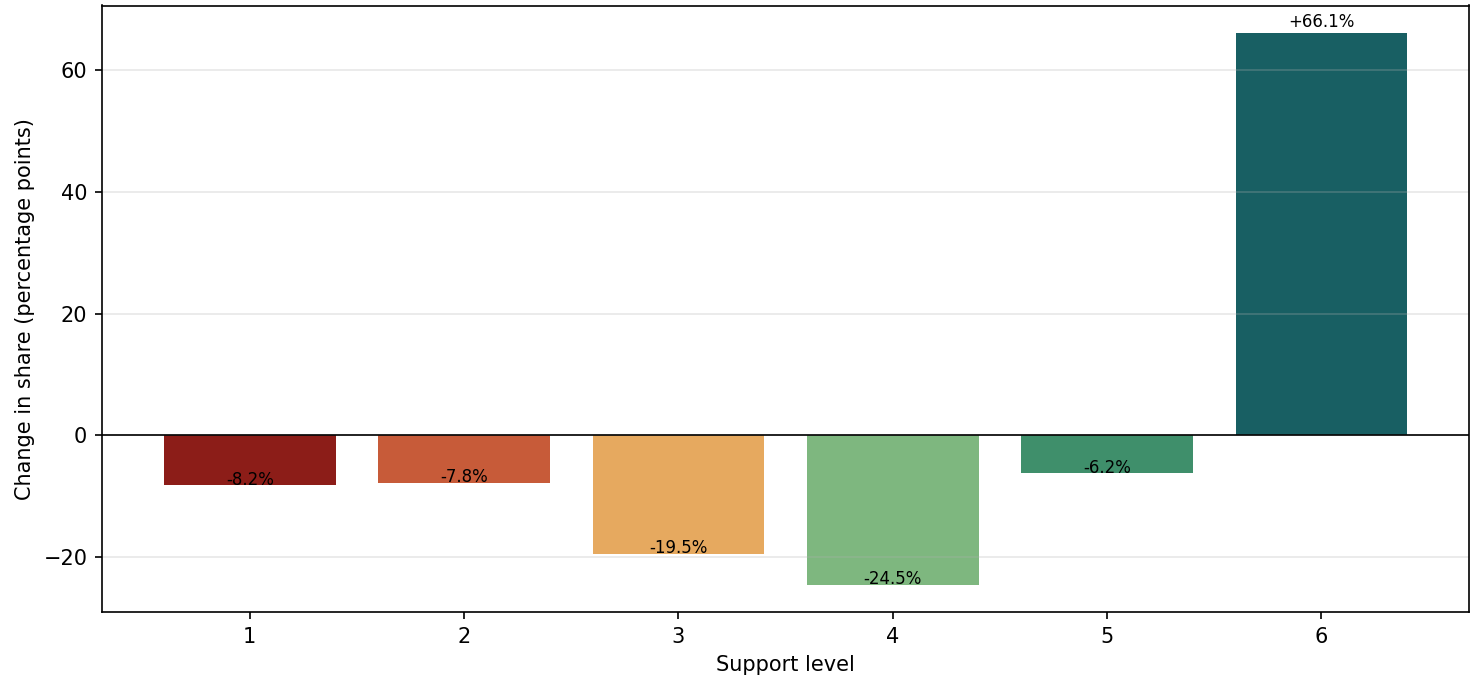}
    \caption{Base scenario, prompt in German, neutral summary.}
    \label{fig:BaseSupportDis2}
\end{subfigure}
\hfill
\begin{subfigure}{0.48\linewidth}
    \centering
    \includegraphics[width=\linewidth]{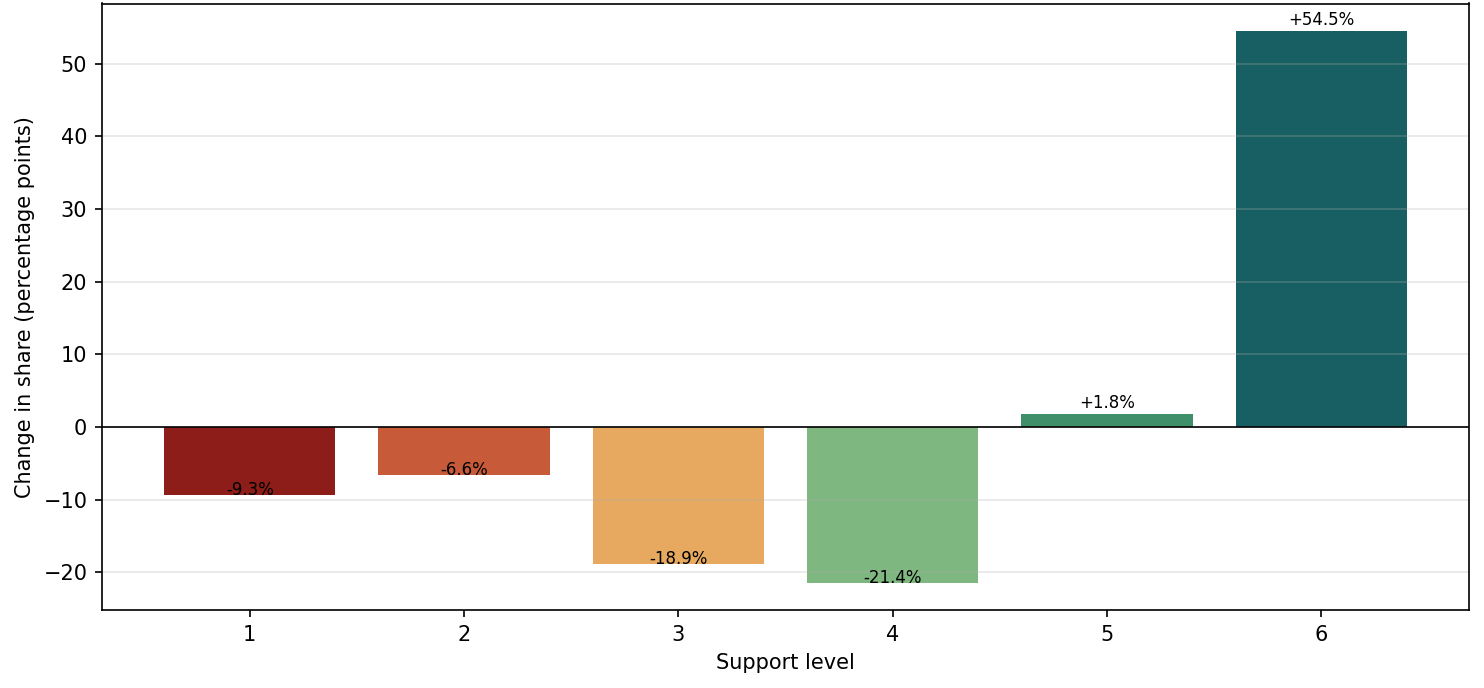}
    \caption{Base scenario, prompt in English, environmental summary.}
    \label{fig:BaseSupportDis3}
\end{subfigure}
\hfill
\begin{subfigure}{0.48\linewidth}
    \centering
    \includegraphics[width=\linewidth]{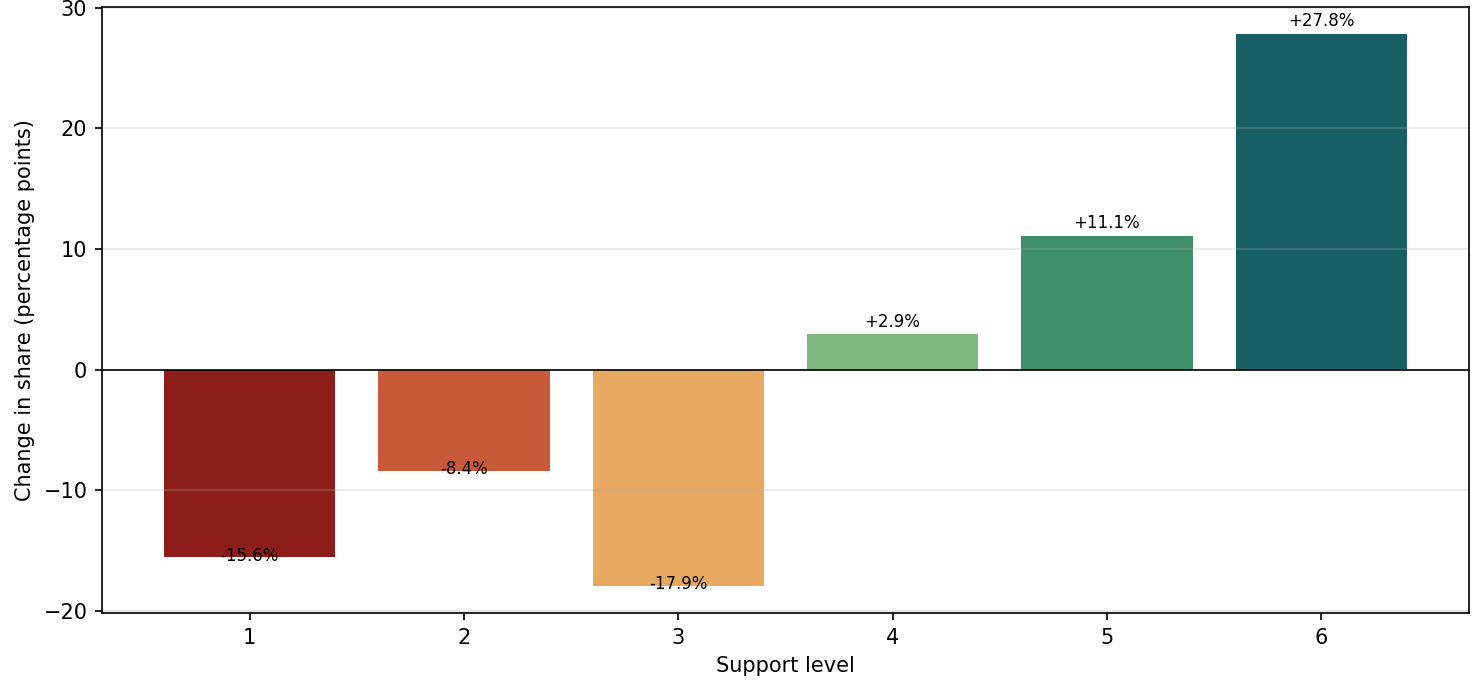}
    \caption{Base scenario, prompt in English, neutral summary.}
    \label{fig:BaseSupportDis4}
\end{subfigure}
\hfill

\begin{subfigure}{0.48\linewidth}
    \centering
    \includegraphics[width=\linewidth]{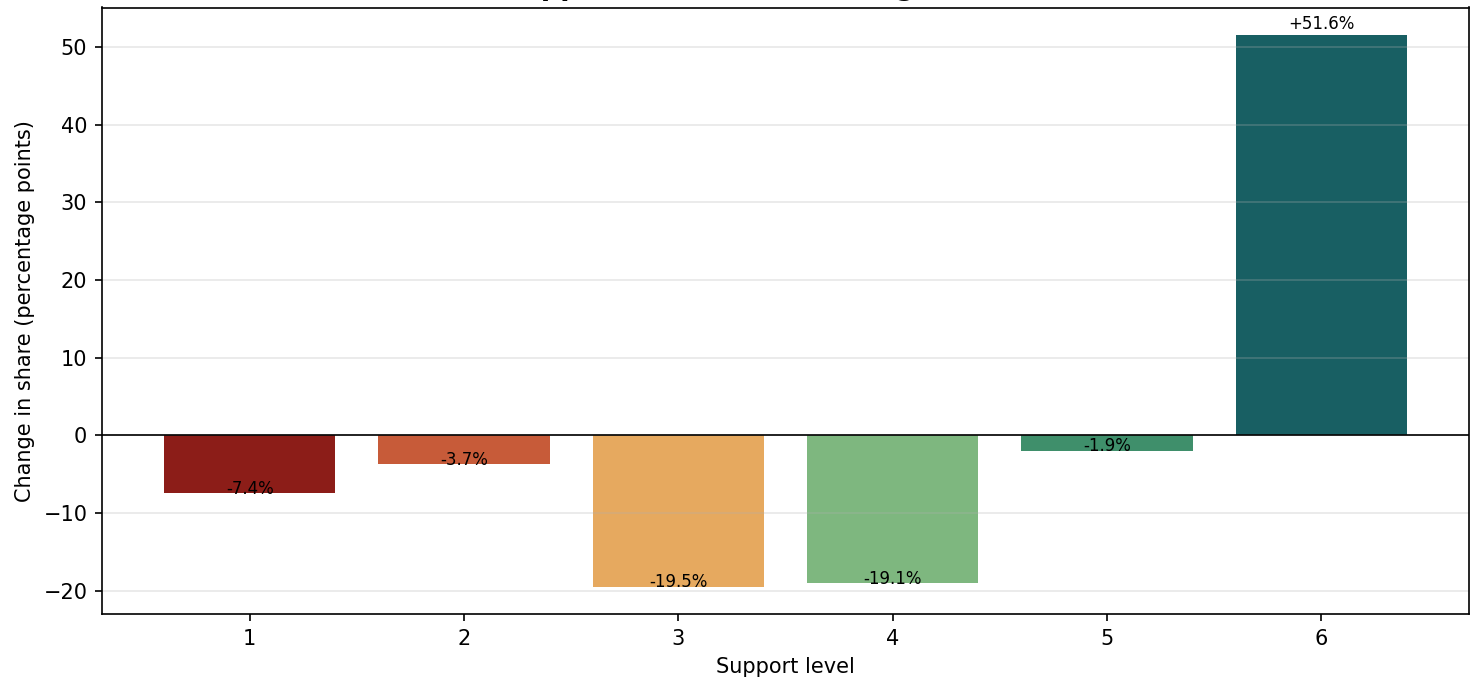}
    \caption{Event scenario, prompt in German, environmental summary.}
    \label{fig:EventSupportDis1}
\end{subfigure}
\hfill
\begin{subfigure}{0.48\linewidth}
    \centering
    \includegraphics[width=\linewidth]{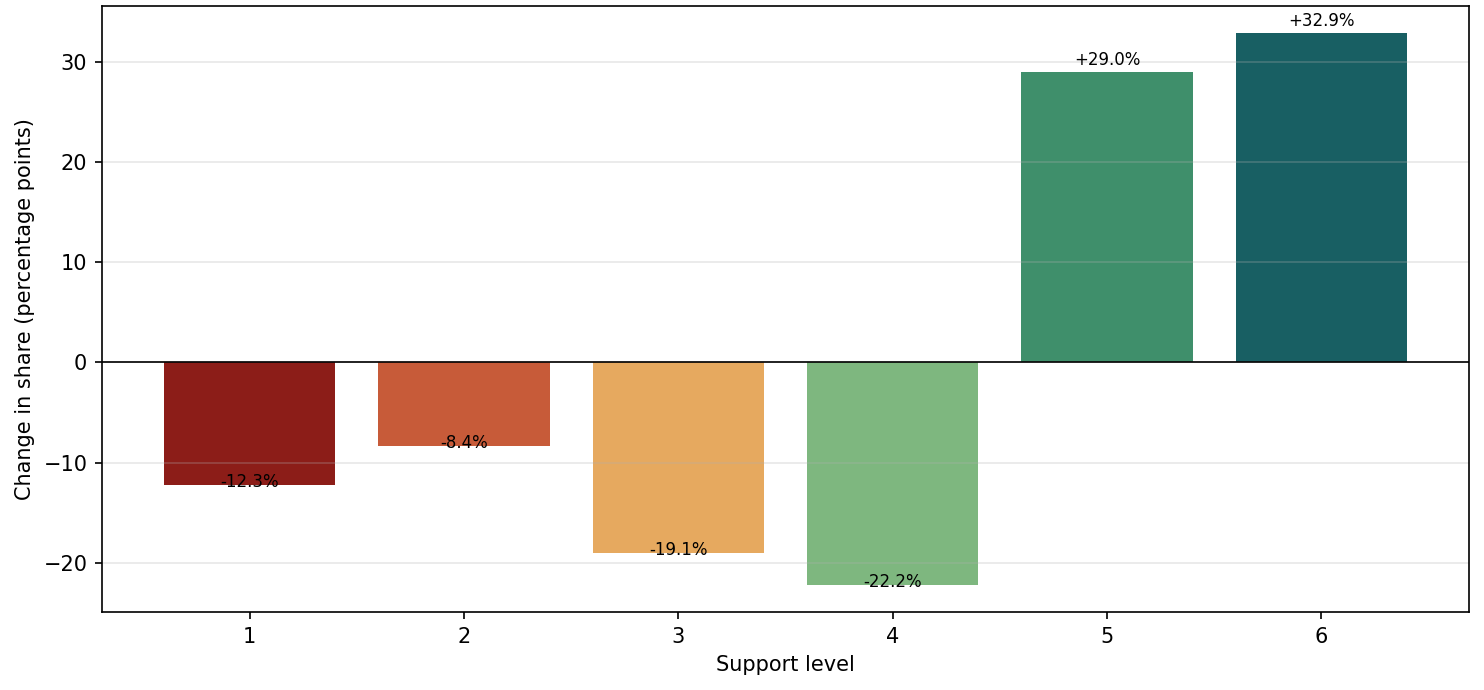}
    \caption{Event scenario, prompt in German, neutral summary.}
    \label{fig:EventSupportDis2}
\end{subfigure}
\hfill
\begin{subfigure}{0.48\linewidth}
    \centering
    \includegraphics[width=\linewidth]{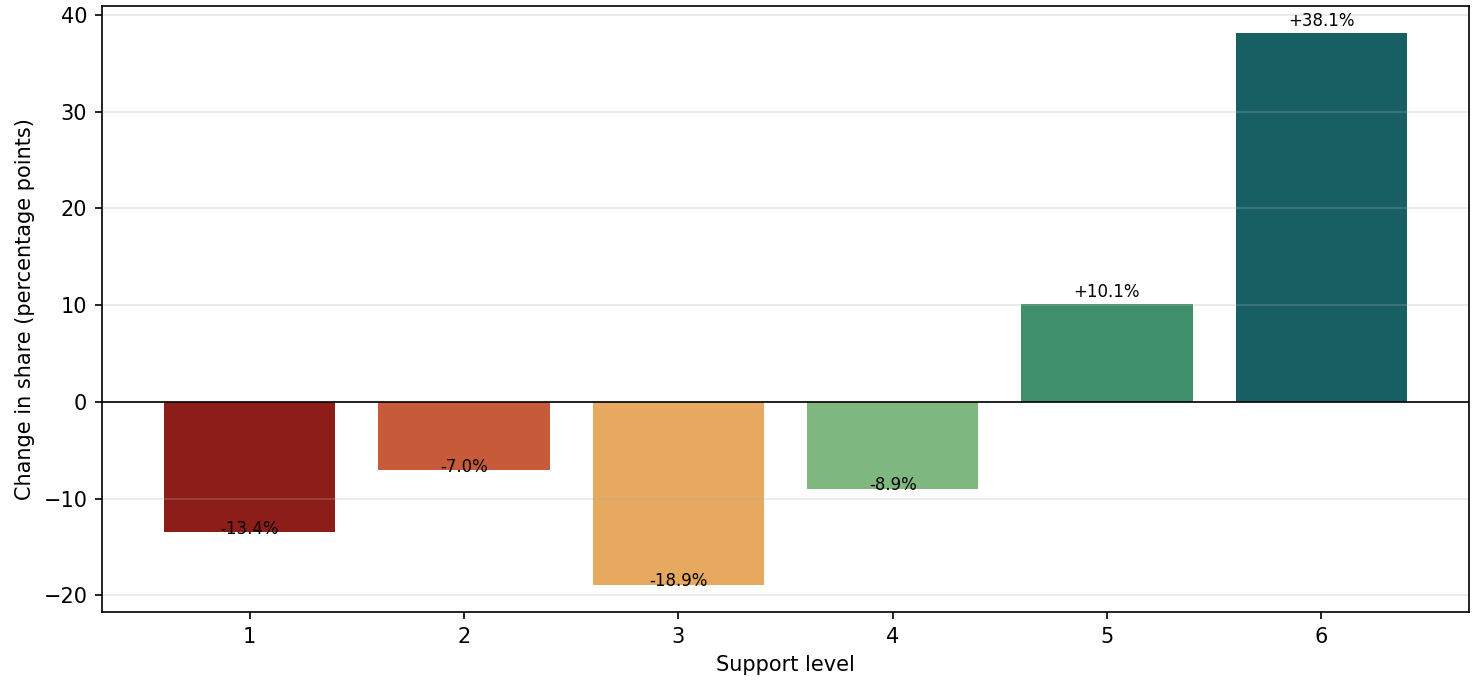}
    \caption{Event scenario, prompt in English, environmental summary.}
    \label{fig:EventSupportDis3}
\end{subfigure}
\hfill
\begin{subfigure}{0.48\linewidth}
    \centering
    \includegraphics[width=\linewidth]{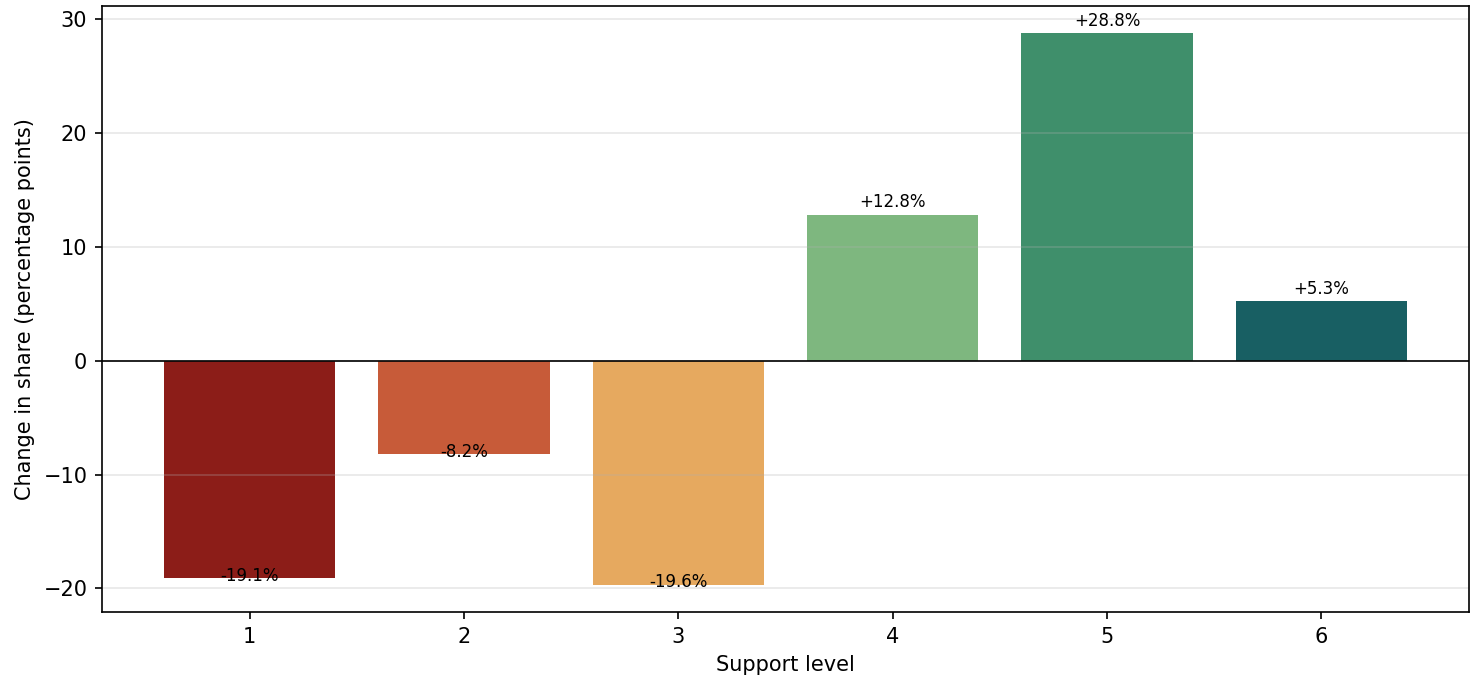}
    \caption{Event scenario, prompt in English, neutral summary.}
    \label{fig:EventSupportDis4}
\end{subfigure}
\hfill
\caption{Support distribution.}
\label{fig:SupportDis}
\end{figure}

In this context, Figure \ref{fig:SupportDis} illustrates the distribution of support across different levels under base and event-driven scenarios, in both prompt languages and styles. In all cases, we observe a notable increase in the share of the highest support level (6). These results indicate that agents eventually diffuse into the optimum support level through what peer agents in their state or national levels contemplate regarding the phase-out. 

In Figures \ref{fig:BaseSupportDis1} and \ref{fig:BaseSupportDis2}, we observe that agents in the last simulation round express very strong support for the phase-out. The range of the support distribution increases from environmental summary to neutral one by over 10\% in the level of 6. When we shift to the results of the base scenario with the English language, we observe slightly different results in Figure \ref{fig:BaseSupportDis3}. Unlike the previous two cases, in this test, we observe a minor distribution of the support in level 5. As soon as we change the prompt summary to a neutral format, the support distribution spans between 4, 5, and 6 levels.

\begin{figure}[H]
\centering

\begin{subfigure}{0.48\linewidth}
    \centering
    \includegraphics[width=\linewidth]{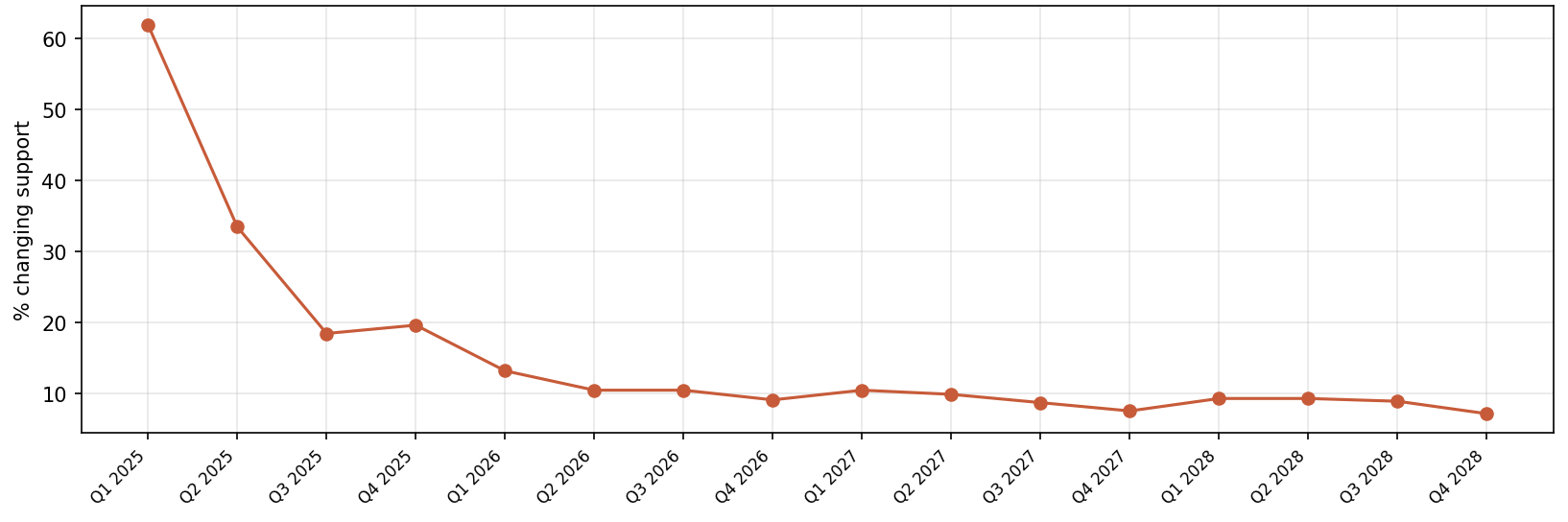}
    \caption{Base scenario, prompt in German, environmental summary.}
    \label{fig:BaseSupportVol1}
\end{subfigure}
\hfill
\begin{subfigure}{0.48\linewidth}
    \centering
    \includegraphics[width=\linewidth]{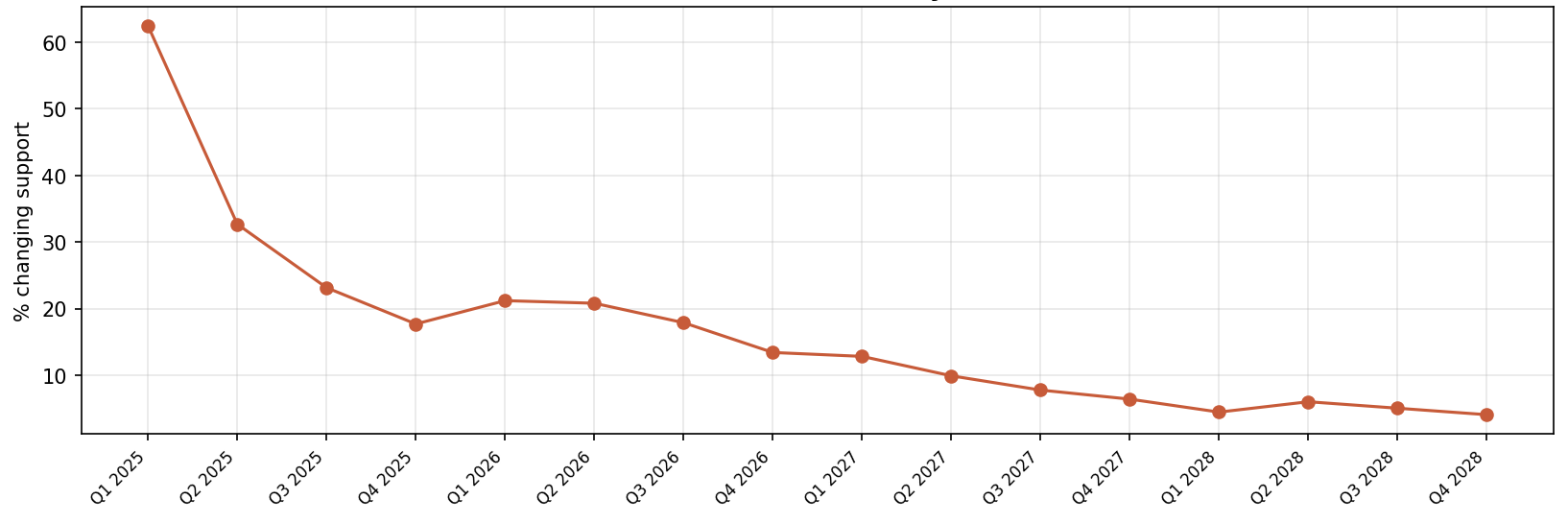}
    \caption{Base scenario, prompt in German, neutral summary.}
    \label{fig:BaseSupportVol2}
\end{subfigure}
\hfill
\begin{subfigure}{0.48\linewidth}
    \centering
    \includegraphics[width=\linewidth]{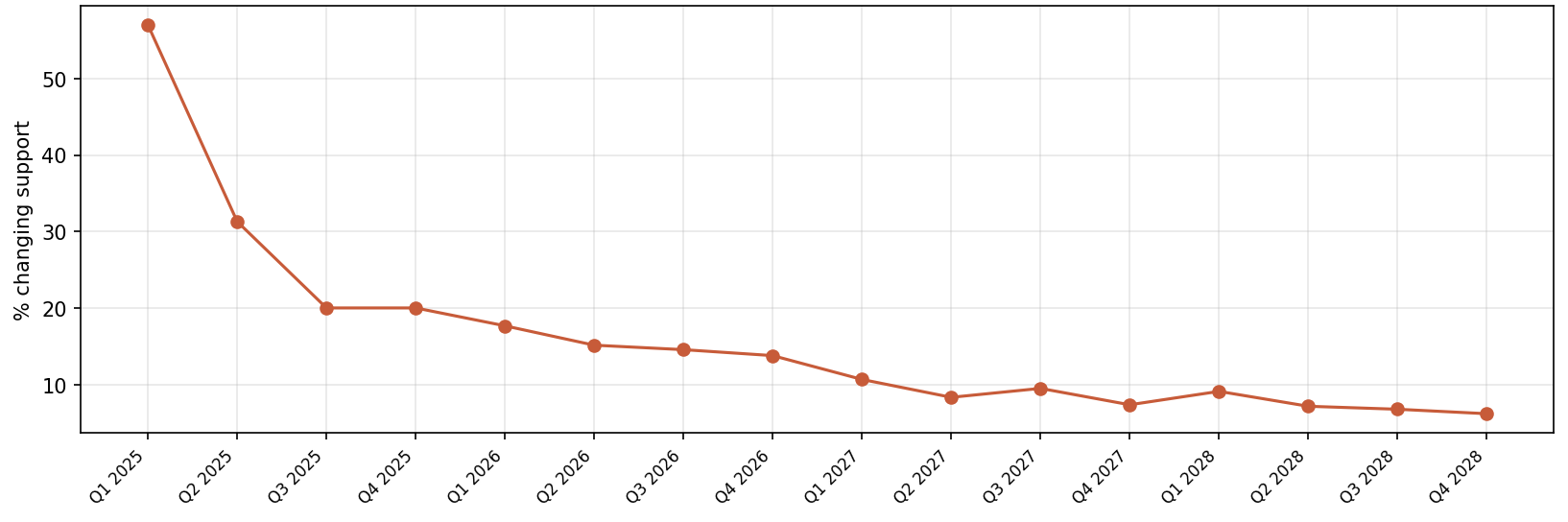}
    \caption{Base scenario, prompt in English, environmental summary.}
    \label{fig:BaseSupportVol3}
\end{subfigure}
\hfill
\begin{subfigure}{0.48\linewidth}
    \centering
    \includegraphics[width=\linewidth]{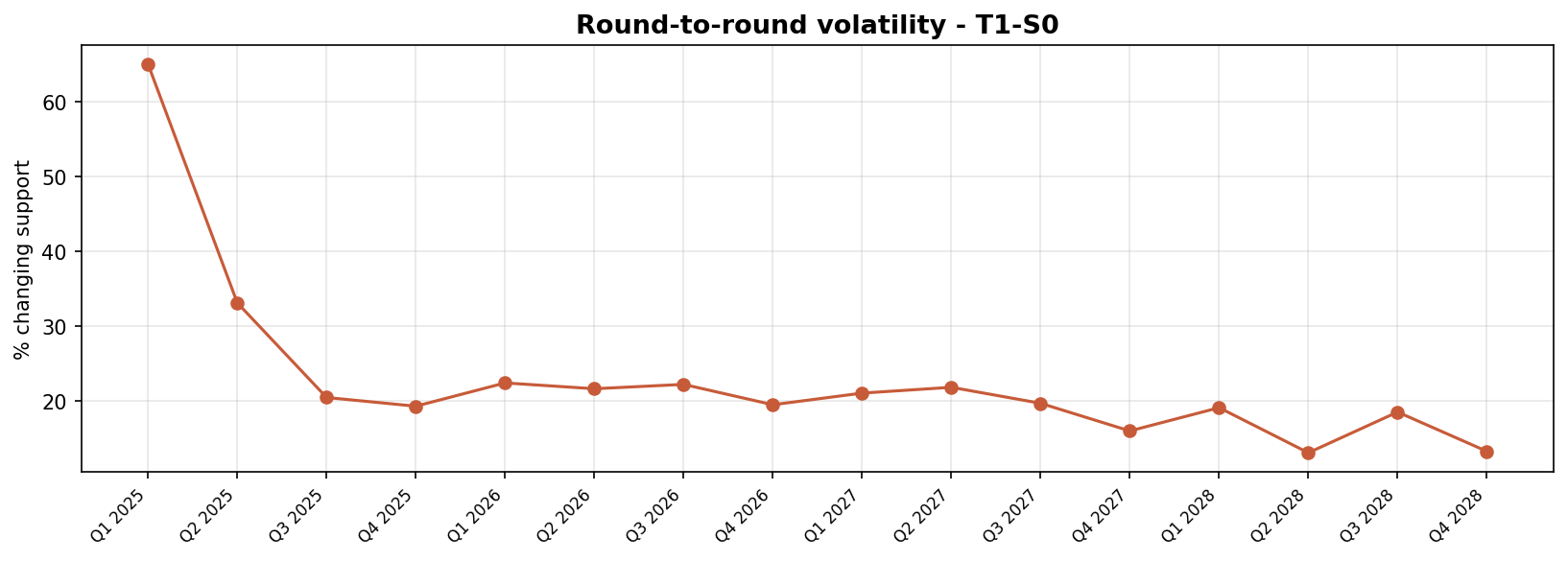}
    \caption{Base scenario, prompt in English, neutral summary.}
    \label{fig:BaseSupportVol4}
\end{subfigure}
\hfill

\begin{subfigure}{0.48\linewidth}
    \centering
    \includegraphics[width=\linewidth]{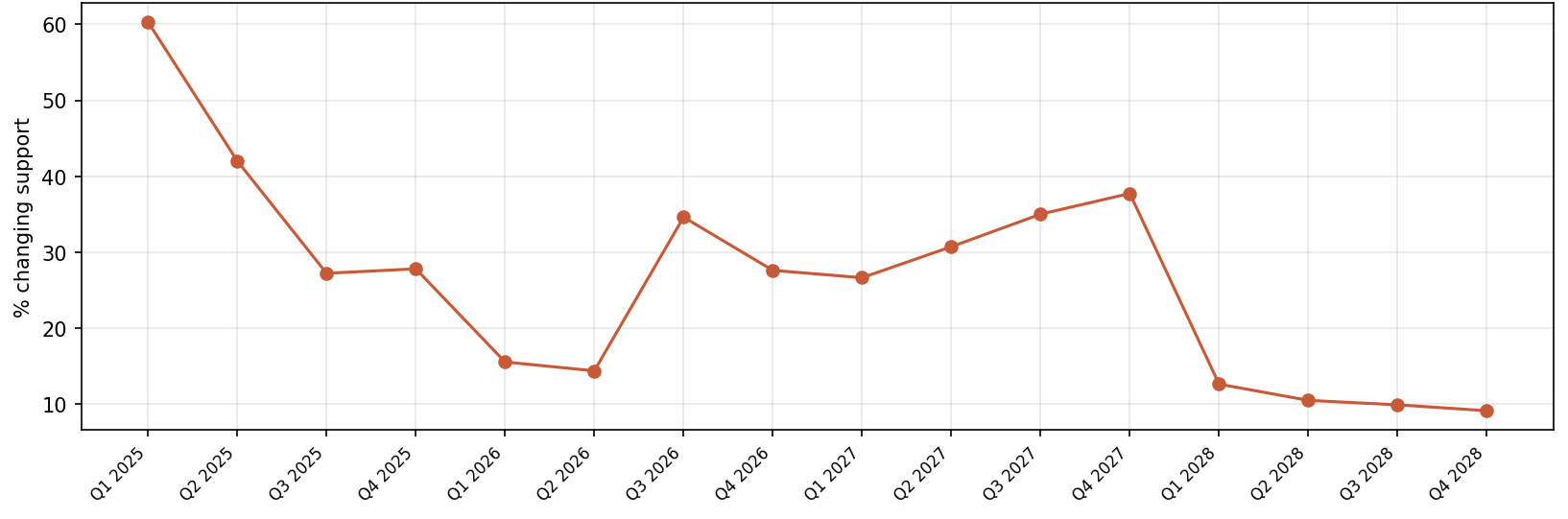}
    \caption{Event scenario, prompt in German, environmental summary.}
    \label{fig:EventSupportVol1}
\end{subfigure}
\hfill
\begin{subfigure}{0.48\linewidth}
    \centering
    \includegraphics[width=\linewidth]{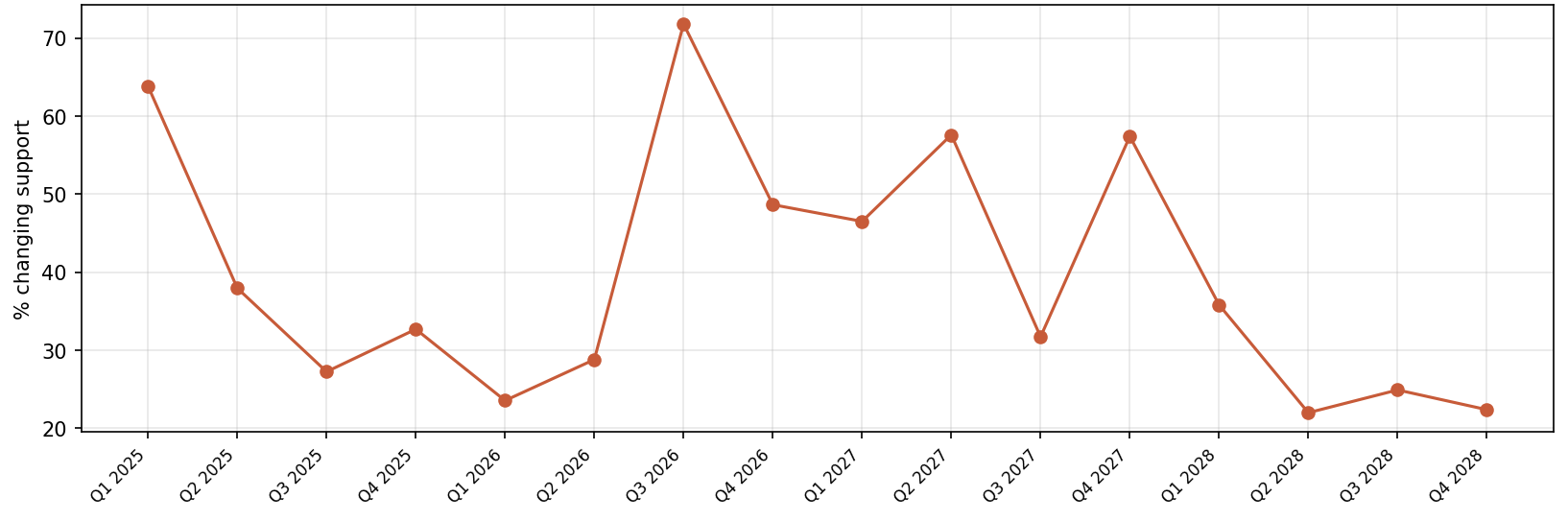}
    \caption{Event scenario, prompt in German, neutral summary.}
    \label{fig:EventSupportVol2}
\end{subfigure}
\hfill
\begin{subfigure}{0.48\linewidth}
    \centering
    \includegraphics[width=\linewidth]{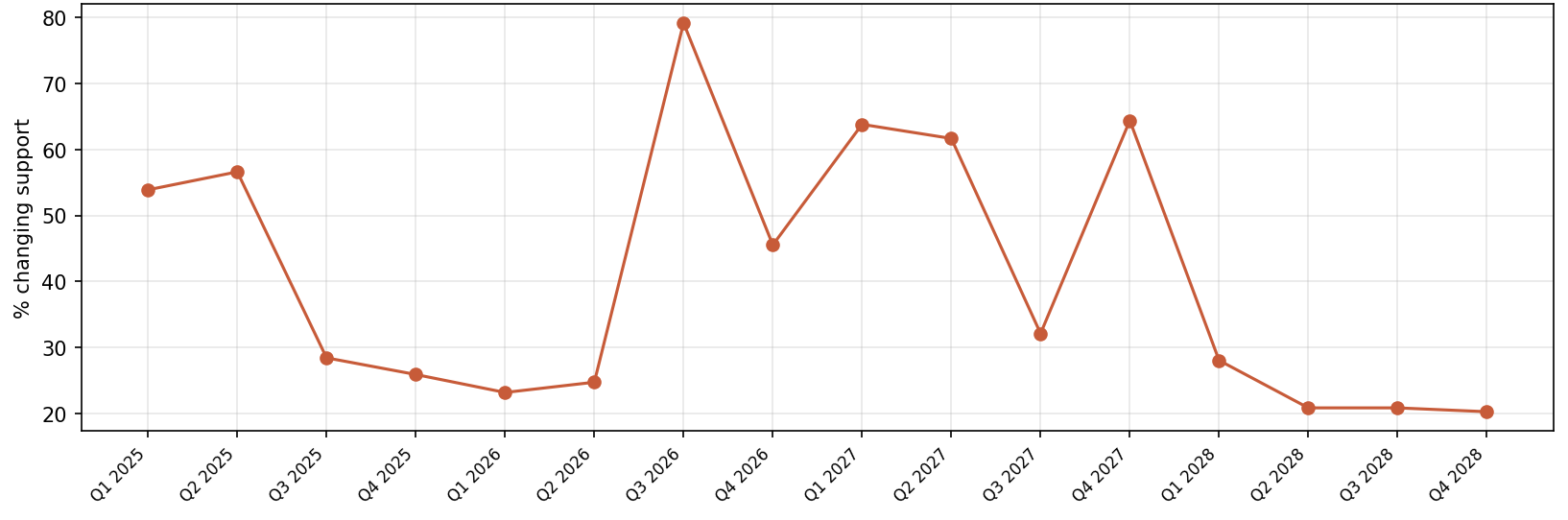}
    \caption{Event scenario, prompt in English, environmental summary.}
    \label{fig:EventSupportVol3}
\end{subfigure}
\hfill
\begin{subfigure}{0.48\linewidth}
    \centering
    \includegraphics[width=\linewidth]{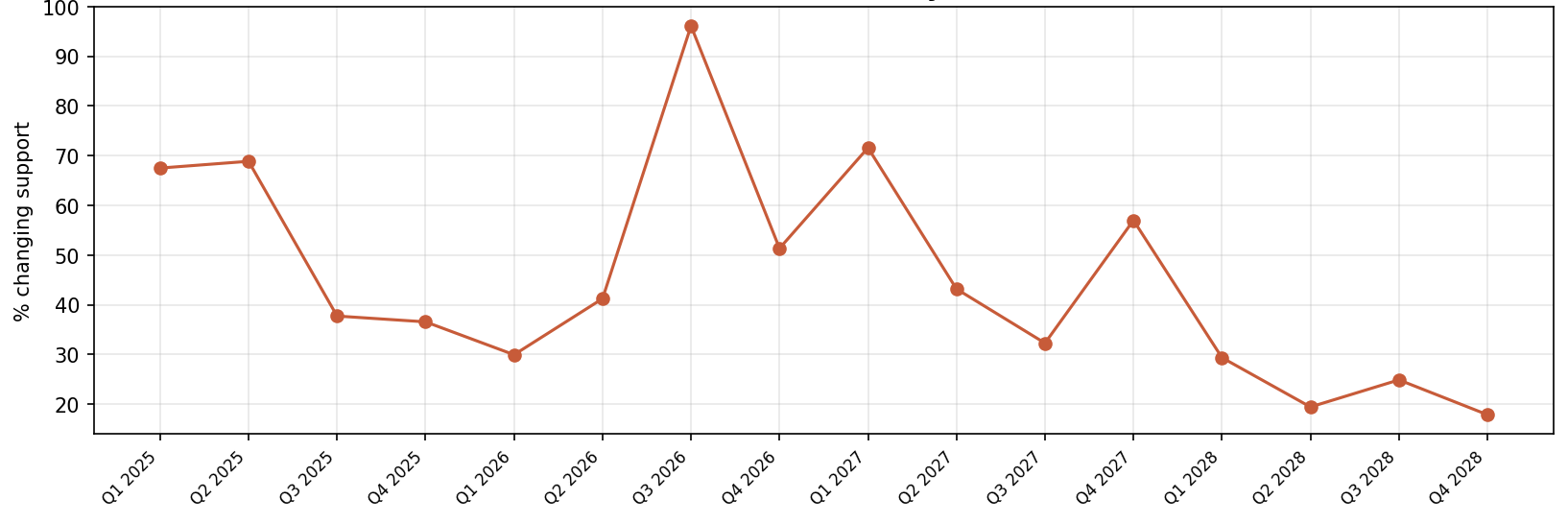}
    \caption{Event scenario, prompt in English, neutral summary.}
    \label{fig:EventSupportVol4}
\end{subfigure}
\hfill
\caption{Support volatility.}
\label{fig:SupportVol}
\end{figure}

In the event-driven scenario with German prompt and environmental summary style, the results demonstrate almost a similar pattern (Figure \ref{fig:EventSupportDis1}). However, relaxing the environmental style from the prompt, we observe a dramatic increase in the share of support in level 5 (Figure \ref{fig:EventSupportDis1}). If we take a look at the same scenario with an English prompt, the most notable trend appears in Figure \ref{fig:EventSupportDis4}, which is based on the neutral summary style. Here, for the first time, results show a higher distribution in level 5 rather than level 6. This highlights the fact that agents without an environmental anchor to the prompt summary show slightly conservative behavior regarding the phase-out. 

To have a better overview of the decisions of the agents, Figure \ref{fig:SupportVol} demonstrates round-to-round switching at the agent-level. For each round, it shows the share of agents whose support level differs from the previous round, so the first point reflects the change from the original survey response to the first simulated decision.

\begin{figure}[H]
\centering

\begin{subfigure}{0.48\linewidth}
    \centering
    \includegraphics[width=\linewidth]{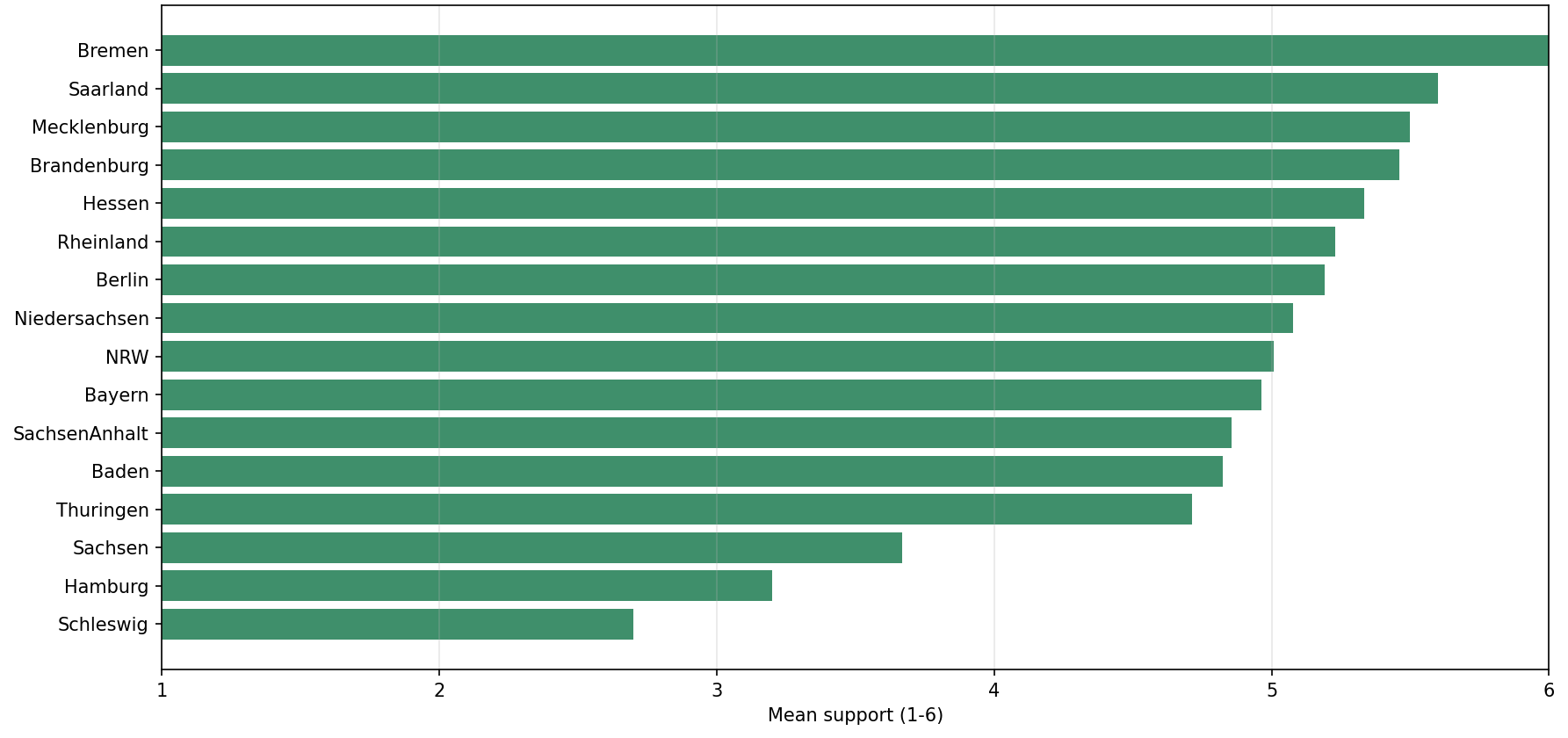}
    \caption{Base scenario, prompt in German, environmental summary.}
    \label{fig:BaseMeanState1}
\end{subfigure}
\hfill
\begin{subfigure}{0.48\linewidth}
    \centering
    \includegraphics[width=\linewidth]{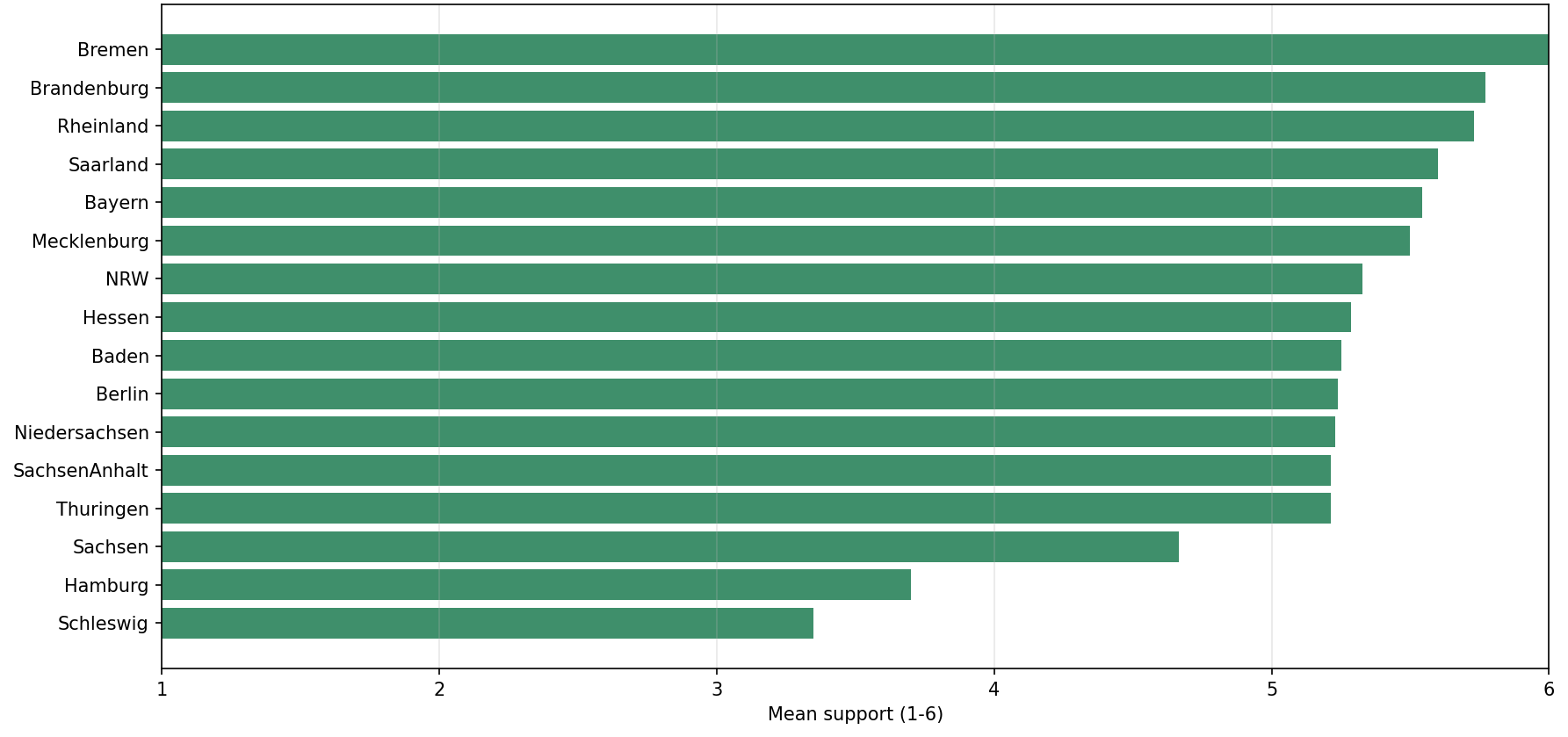}
    \caption{Base scenario, prompt in German, neutral summary.}
    \label{fig:BaseMeanState2}
\end{subfigure}
\hfill
\begin{subfigure}{0.48\linewidth}
    \centering
    \includegraphics[width=\linewidth]{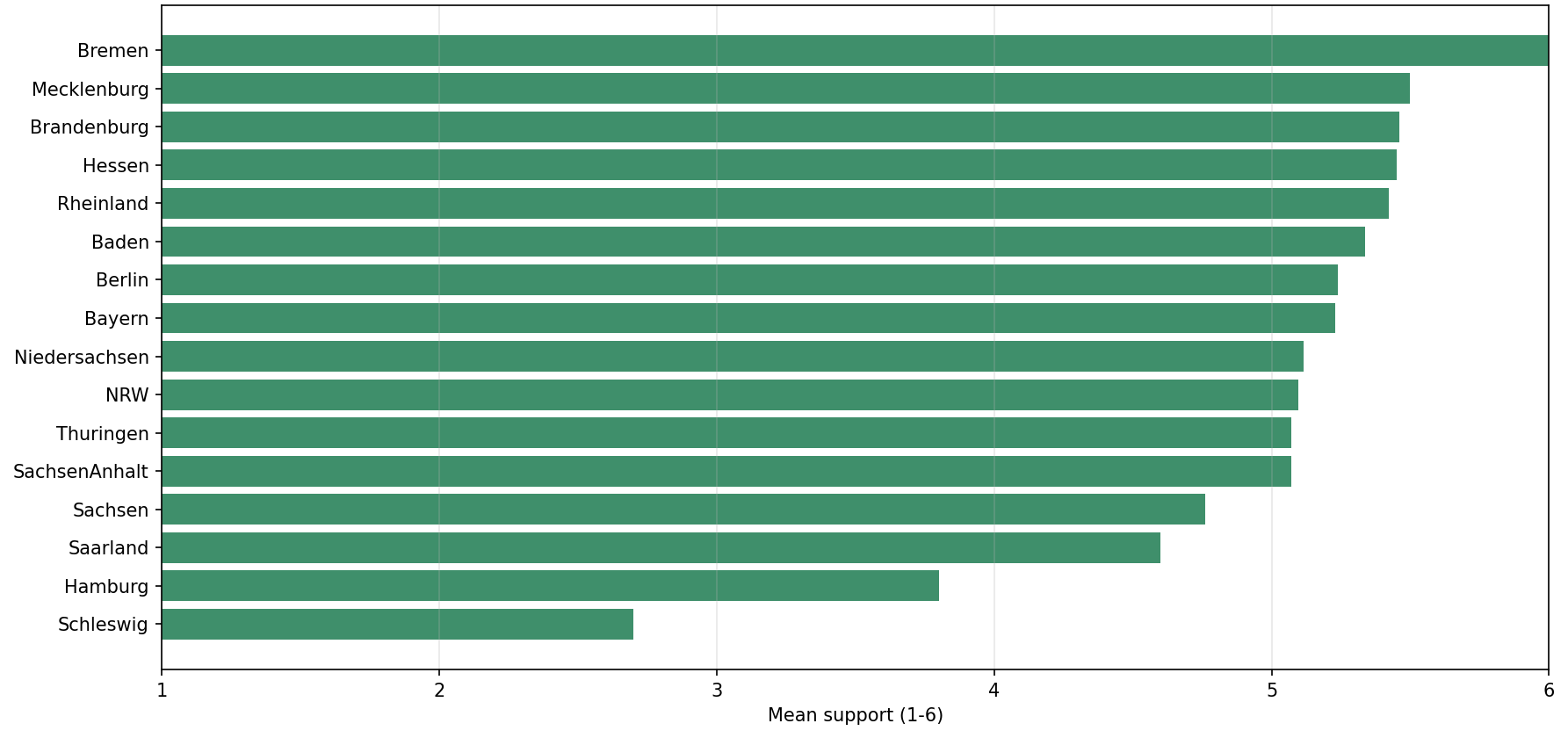}
    \caption{Base scenario, prompt in English, environmental summary.}
    \label{fig:BaseMeanState3}
\end{subfigure}
\hfill
\begin{subfigure}{0.48\linewidth}
    \centering
    \includegraphics[width=\linewidth]{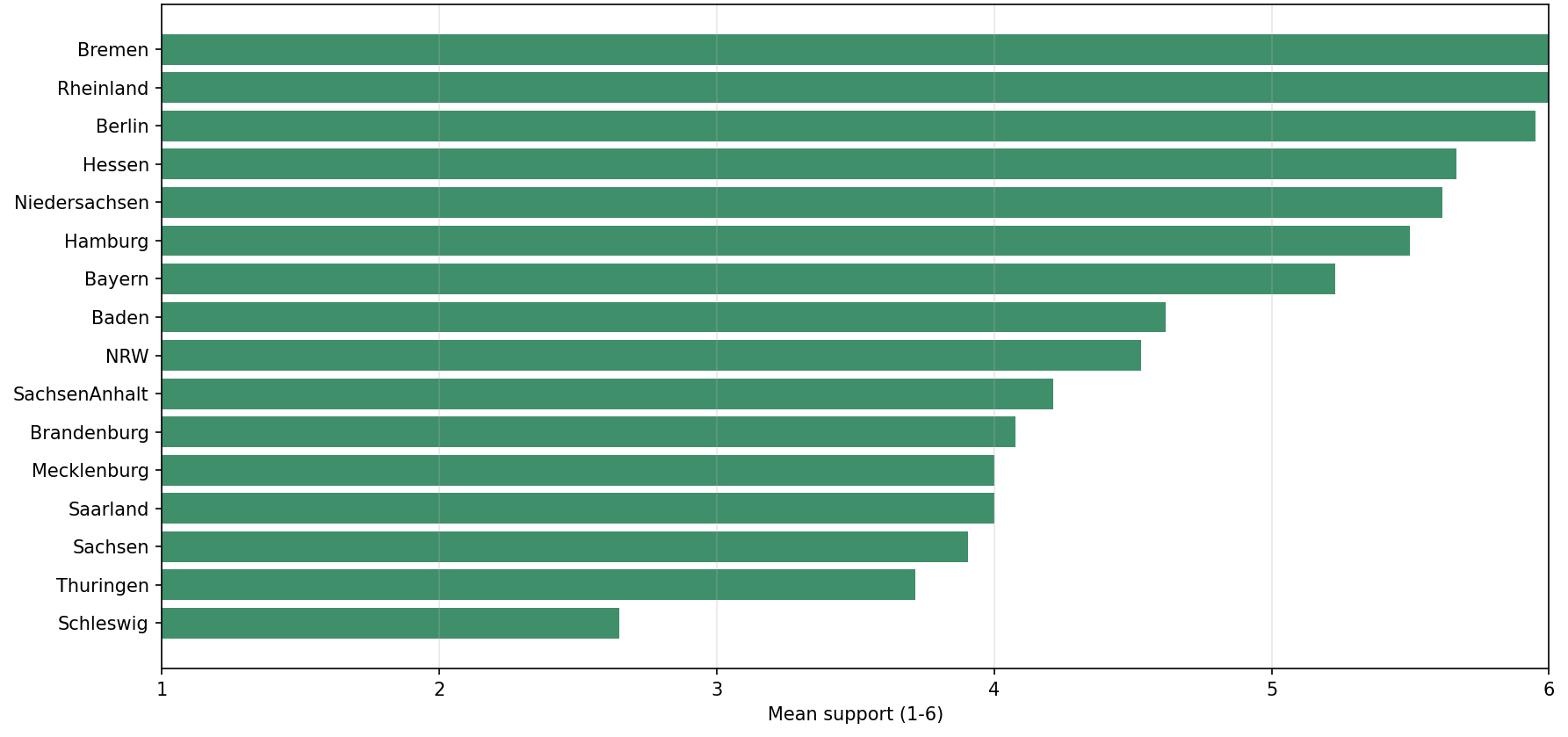}
    \caption{Base scenario, prompt in English, neutral summary.}
    \label{fig:BaseMeanState4}
\end{subfigure}
\hfill

\begin{subfigure}{0.48\linewidth}
    \centering
    \includegraphics[width=\linewidth]{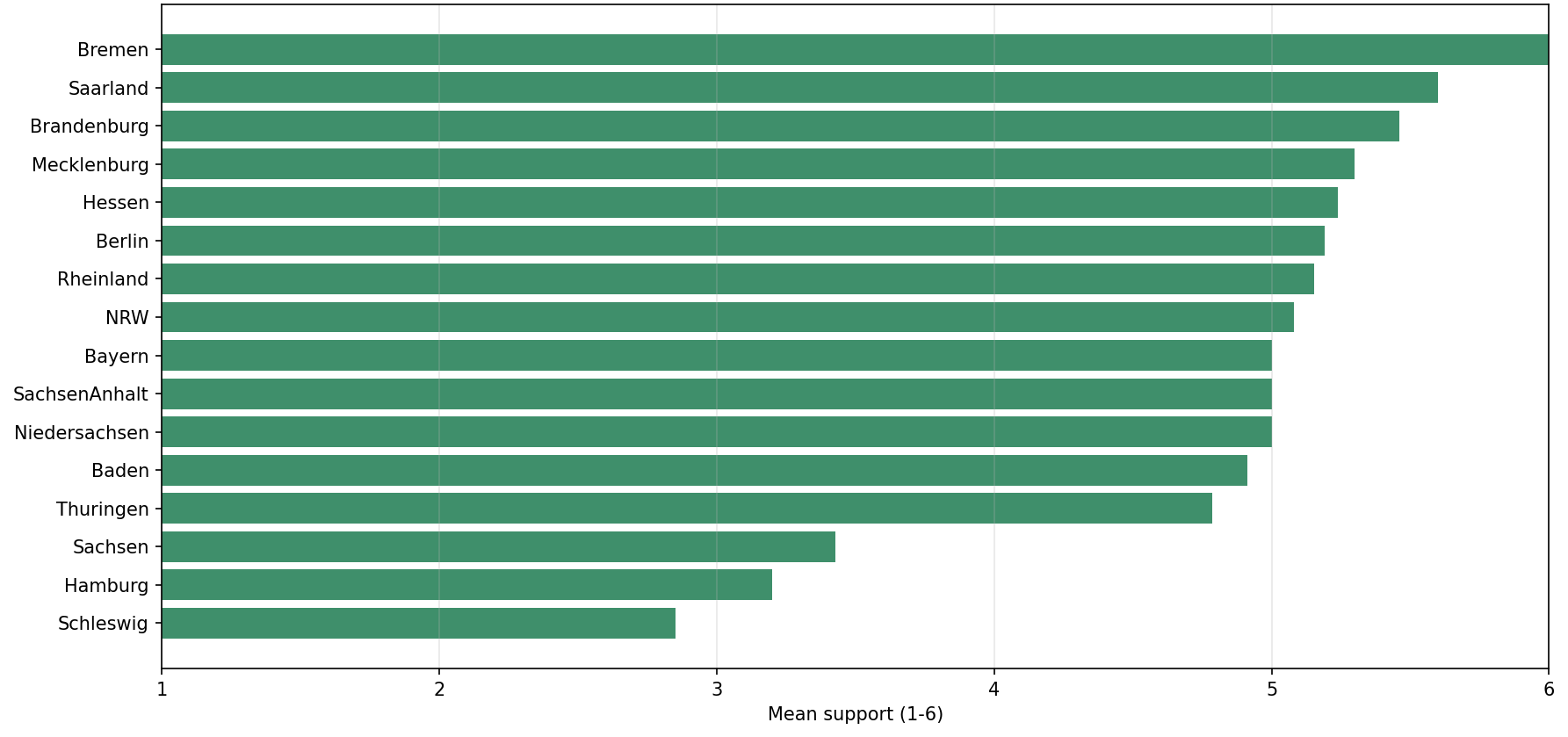}
    \caption{Event scenario, prompt in German, environmental summary.}
    \label{fig:EventMeanState1}
\end{subfigure}
\hfill
\begin{subfigure}{0.48\linewidth}
    \centering
    \includegraphics[width=\linewidth]{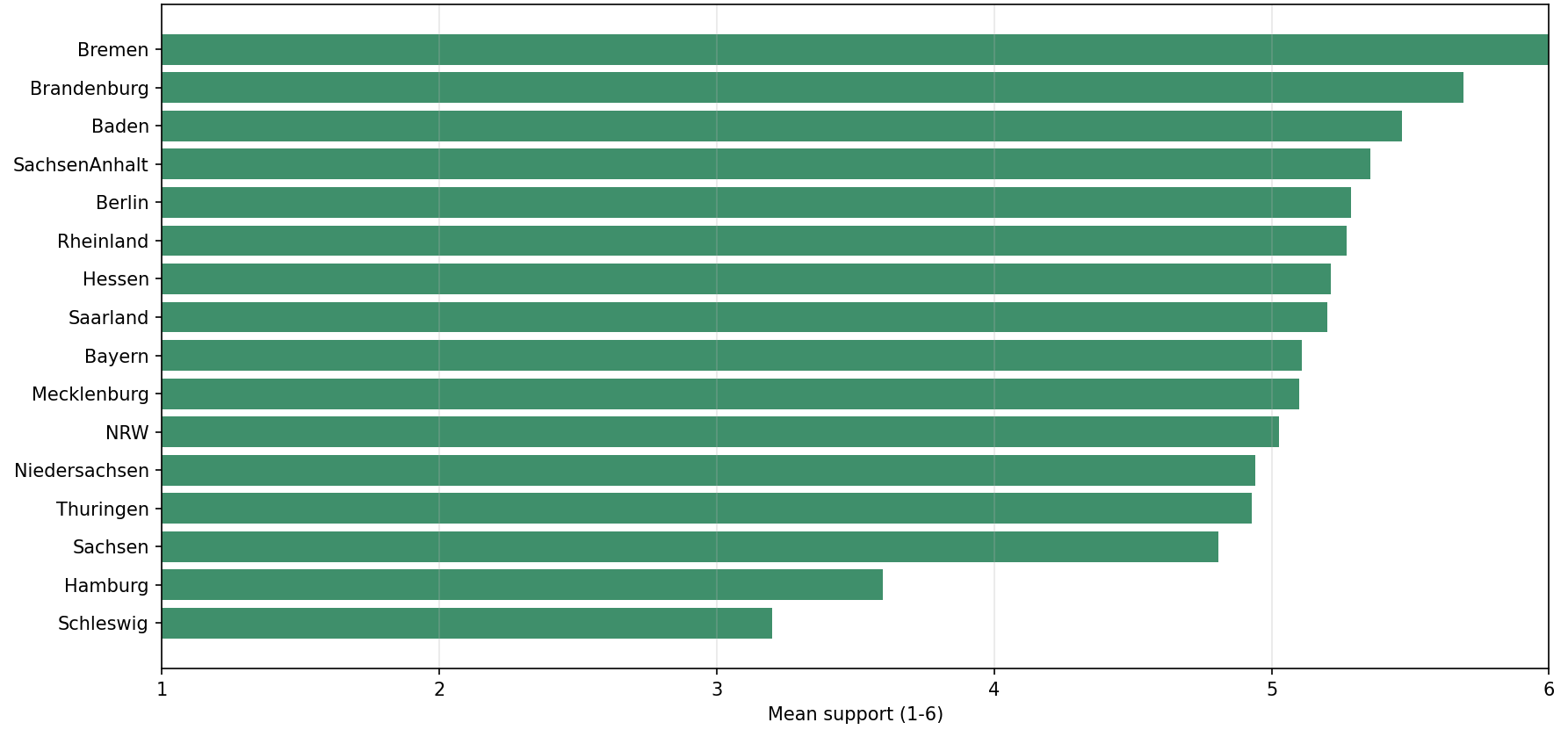}
    \caption{Event scenario, prompt in German, neutral summary.}
    \label{fig:EventMeanState2}
\end{subfigure}
\hfill
\begin{subfigure}{0.48\linewidth}
    \centering
    \includegraphics[width=\linewidth]{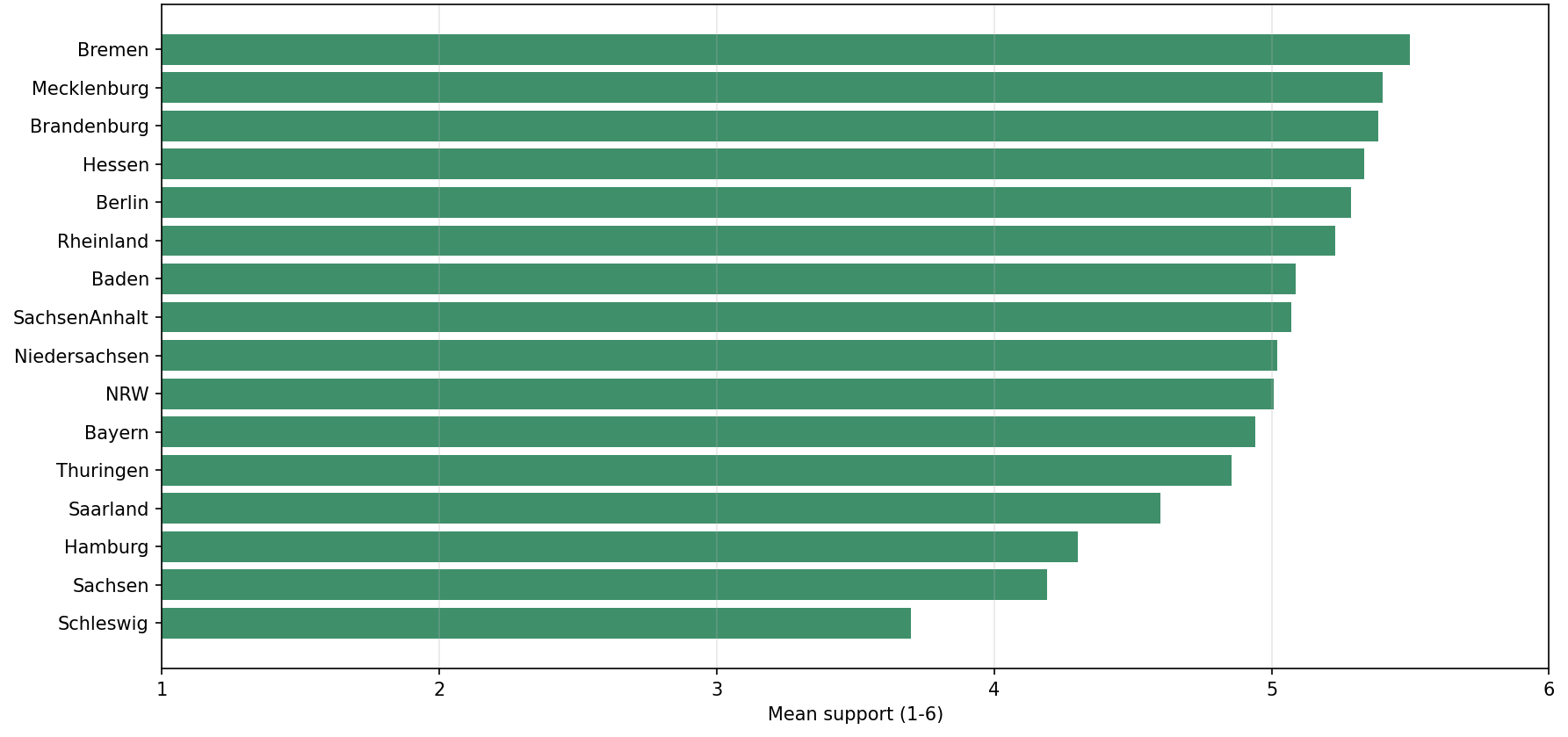}
    \caption{Event scenario, prompt in English, environmental summary.}
    \label{fig:EventMeanState3}
\end{subfigure}
\hfill
\begin{subfigure}{0.48\linewidth}
    \centering
    \includegraphics[width=\linewidth]{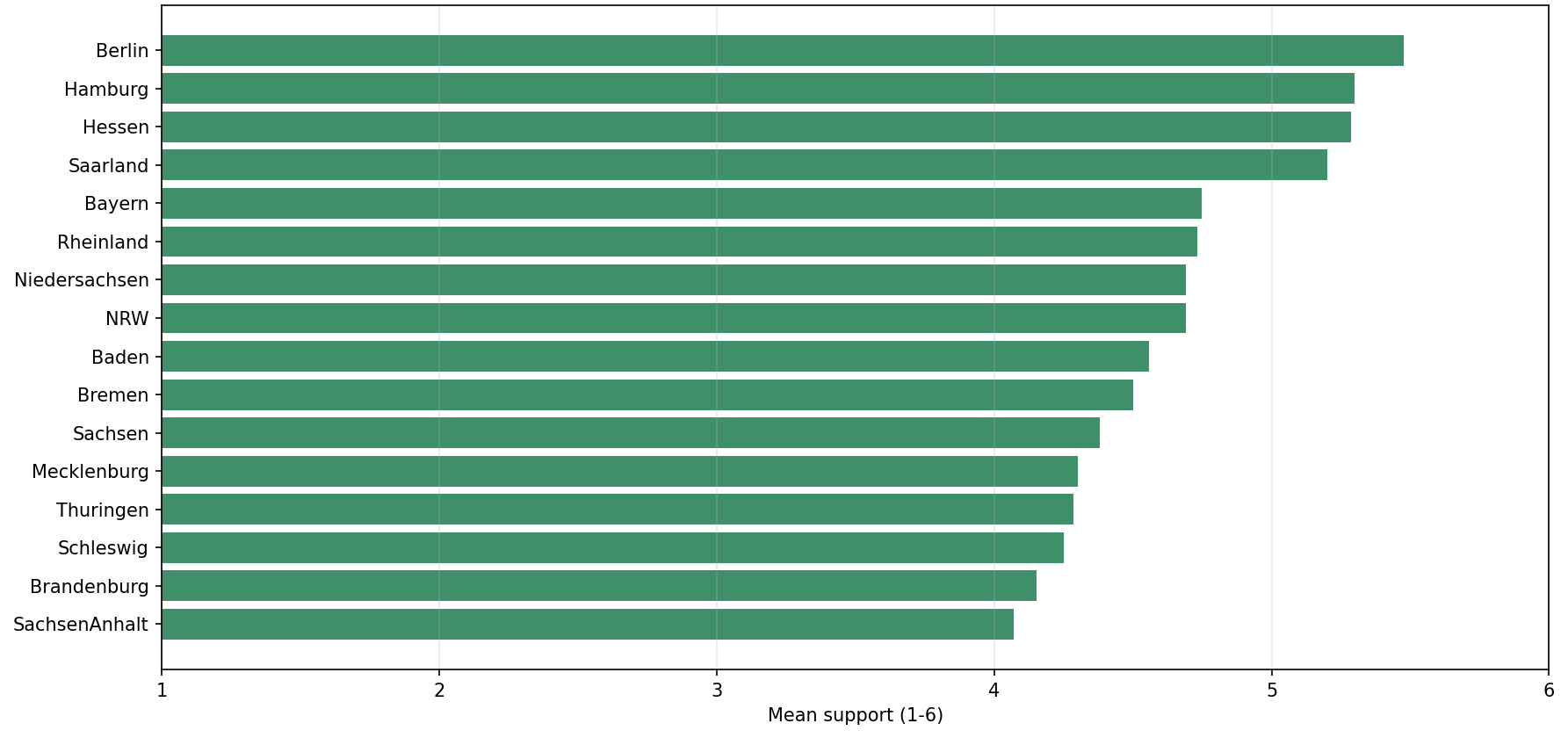}
    \caption{Event scenario, prompt in English, neutral summary.}
    \label{fig:EventMeanState4}
\end{subfigure}
\hfill
\caption{Mean support in federal states in the last simulation round.}
\label{fig:MeanState}
\end{figure}

Finally, it is important to understand how agents at the state level perform over time. This is crucial because national averages can disguise substantial regional heterogeneity in policy preferences. Examining state-level trajectories, therefore, allows us to see whether the simulation captures sub-national variation and whether the effects of events differ across the federal states. Figure \ref{fig:MeanState} presents an overview of the support level across federal states in the last simulation round. 

In almost all of the simulation cases, agents in Bremen demonstrate the highest level of support for voting for the parties, which includes the ICE phase-out as part of their political plan. Only in the simulation with events, prompted in English with neutral summary style, agents in Berlin present higher support compared to others (Figure \ref{fig:EventMeanState4}). Additionally, Rheinland-Pfalz is the second federal state to show the highest level of support (6) in Figure \ref{fig:BaseMeanState2}, which is a base scenario case with the prompt in English and neutral summary style. 

On the contrary, agents in Schleswig-Holstein state have shown the lowest level of support to the ICE phase-out policy in almost all simulation cases (Figure \ref{fig:MeanState}). Sachsen-Anhalt (Saxony-Anhalt) is the only state to show a lower support level compared to Schleswig-Holstein, in the simulation with events, prompted in English with a neutral summary style (Figure \ref{fig:EventMeanState4}). Schleswig-Holstein is also the only state with a support level below (3) (Figures \ref{fig:BaseMeanState1}, \ref{fig:BaseMeanState3}, \ref{fig:BaseMeanState4}, \ref{fig:EventMeanState1}). Further insights could be derived based on Figure \ref{fig:MeanState}.

\end{document}